\documentclass[11pt]{article}

\usepackage{amsmath}
\usepackage{amsfonts}
\usepackage{amsthm}
\usepackage{amssymb, xcolor}
\usepackage[numbers]{natbib}
\usepackage{fullpage}
\usepackage{times}

\newcommand{\ie}{{\it i.e.}}
\newcommand{\BEAS}{\begin{eqnarray*}}
\newcommand{\EEAS}{\end{eqnarray*}}
\newcommand{\BEA}{\begin{eqnarray}}
\newcommand{\EEA}{\end{eqnarray}}
\newcommand{\BIT}{\begin{itemize}}
\newcommand{\EIT}{\end{itemize}}
\newcommand{\BEQ}{\begin{equation}}
\newcommand{\EEQ}{\end{equation}}
\newcommand{\BNUM}{\begin{enumerate}}
\newcommand{\ENUM}{\end{enumerate}}

\newcommand{\Mv}{(v_1, \ldots, v_n)}
\newcommand{\M}{\mathcal{M}}

\newcommand{\at}{q^*}

\newcommand{\reals}{{\mathbb{R}}}
\newcommand{\naturals}{{\mathbb{N}}}
\newcommand{\rvm}{{m}}
\newcommand{\strat}{{\pi}}
\newcommand{\spart}{\mathcal{S}}
\newcommand{\qpart}{{\bf q}}

\newcommand{\estrat}{{\strat^E}}
\newcommand{\sstrat}{{\strat^S}}
\newcommand{\sstrati}{\strat^S_1}
\newcommand{\sstratj}{\strat^S_2}
\newcommand{\eopt}{\mathcal{E}^*}
\newcommand{\spartj}{\spart(\sstratj)}%Set of participants in WTA in another semi-equilibrium (reqd for second lemma)
\newcommand{\opt}{\mathrm{OPT}}
\newcommand{\wta}{\mathrm{WTA}}
\newcommand{\WTA}{\mathcal M_{\mathrm{WTA}}}
\newcommand{\vc}{{\nu}}
\newcommand{\ld}{{\lambda}}

\DeclareMathOperator{\argmax}{arg\,max}

\newcommand{\tr}{\textcolor{red}}

\newcommand{\ntb}{}

\newtheorem{definition}{Definition}[section]
\newtheorem{example}{Example}[section]
\newtheorem{lemma}{Lemma}[section]
\newtheorem{proposition}{Proposition}[section]
\newtheorem{theorem}{Theorem}[section]
\newtheorem{corollary}{Corollary}[section]

\begin{document}
\title{Behavioral Mechanism Design:\\
Optimal Contests for Simple Agents}
\author{Arpita Ghosh and Robert Kleinberg\\
Cornell University\\
Ithaca, NY, USA}
\date{}

\maketitle

\begin{abstract}
Incentives are more likely to elicit desired outcomes when they are designed based on accurate models of agents' strategic behavior.
%
%\textcolor{green}{Economic design is more likely to be effective when it is accurately models the agents it is targeted at.}
%
A growing literature, however, suggests that people do not quite behave like standard economic agents in a variety of environments, both online and offline. What consequences might such differences have for the optimal {\em design} of mechanisms in these environments? 
In this paper, we explore this question in the context of optimal contest design for {\em simple} agents---agents who strategically reason about whether or not to participate in a system, but not about the input they provide to it. 
Specifically, consider a contest where $n$ potential contestants with types $(q_i,c_i)$ each choose between participating and producing a submission of quality $q_i$ at cost $c_i$, versus not participating at all, to maximize their utilities. How should a principal distribute a total prize $V$ amongst the $n$ ranks to maximize some increasing function of the qualities of elicited submissions in a contest with such simple agents?

We first solve the optimal contest design problem for settings where agents have homogenous participation costs $c_i = c$. Here, the contest that maximizes every increasing function of the elicited contributions is always a {\em simple} contest, awarding equal prizes of $V/j^*$ each to the top $j^*
=V/c - \Theta(\sqrt{V/(c\ln(V/c))})$ contestants. 
 %= \vc - \Theta(\sqrt{\vc/\ln(\vc)})$ contestants, where $\vc=V/c$. 
This is in contrast with the optimal contest structure in comparable models with strategic effort choices, where the optimal contest is either a winner-take-all contest or awards possibly unequal prizes, depending on the curvature of agents' effort cost functions. We next address the general case with heterogenous costs where agents' types $(q_i,c_i)$ are inherently two-dimensional, significantly complicating equilibrium analysis. With heterogenous costs, the optimal contest depends on the objective being maximized: our main result here is that the winner-take-all contest is a 3-approximation of the optimal contest when the principal's objective is to maximize the quality of the best elicited contribution. The proof of this result hinges around a `sub-equilibrium' lemma establishing a stochastic dominance relation between the distribution of qualities elicited in an equilibrium and a {\em sub-equilibrium}---a strategy profile that is a best response for all agents who choose to {\em participate} in that strategy profile; this relation between equilibria and sub-equilibria may be of more general interest.
\end{abstract}

\maketitle

\section{Introduction} 
\label{s-intro}
A vast range of online systems---social computing and crowdsourcing platforms, matching markets, ranking and recommendation systems---involve agents who make choices that determine the input to, and therefore the outcome of, the system. This poses an immediate problem of designing incentives and environments that drive choices resulting in desirable outcomes, and has resulted in a rich literature on mechanism design and algorithmic game theory for these settings.

%Economic design is more likely to be effective when it accurately models the agents at which it is targeted. 
Economic design is more likely to be effective when it is based on accurate models of the agent population at which it is targeted. 
A growing literature, however, suggests that people do not quite behave like standard economic agents in a variety of settings, both online and offline. What consequences might such differences in behavior have for the optimal {\em design} of these environments? 

In this paper, we explore this question of `behavioral' mechanism design in the context of designing optimal contests for {\em simple} agents---agents who strategically reason about whether or not to participate in a system, but not about the input they provide to it---to see how this difference in agent behavior affects the analysis of the strategic environment and the structure of optimal mechanisms.  

Before proceeding, we note that our use of the term `behavioral' differs slightly from that in the behavioral economics literature, and specifically is not meant to {\em require} that our agents are irrational in any sense or have bounded reasoning abilities: indeed, the simple agents in our model do not make suboptimal choices---they do make optimal choices, but over restricted {\em choice} sets relative to the standard models in the contest design literature. 
We use the term behavioral only to indicate (mechanism design problems with) models of agents that  more accurately capture real human behavior in the environment of interest than standard economic models.
%Our use of the term `behavioral' is only meant to indicate that agents in these mechanism design problems behave differently in some way from a `standard' economic agent, and that the difference aims to capture a difference between standard economic models and more realistic models of human behavior.} 

\subsection{Contest design} 
Contests, where participants expend some resource---time, effort, money--- to compete for prizes, are everywhere. There are athletic and artistic competitions and competitions for mathematics and research development and design; online, too, there an exploding number of contests---both explicit, like online photography contests and design contests on platforms like Quirky and 99designs\footnote{www.quirky.com; www.99designs.com}, as well as implicit contests for attention or virtual rewards\footnote{Most user-generated content sites rank users’ contributions according to some measure of quality, and a contribution’s ranking determines its likely viewership and therefore the attention reward to the corresponding user. A number of websites also give out more explicit virtual rewards to encourage contribution, such as virtual points (e.g. Y! Answers), badges (StackOverflow, Quora, TripAdvisor) or leaderboard rankings (Amazon, Y! Answers).}  on user-contribution based websites. The ubiquitousness of contests, as well as the seminal work of~\citet{Lazear81} showing that contests can serve as an efficient incentive scheme for effort elicitation, has led to a huge literature on contest design, studying how to allocate rewards to best incentivize high-quality submissions from utility-maximizing contestants with a cost to effort.

There is a wide spectrum of contests, however, where participants do not quite strategize over the quality of their submissions: 
\BIT 
\item First, there are contests where the {\em production} of an entry is essentially costless, and the only cost incurred is that of submission---for instance, potential contestants may already possess eligible entries prior to, or independent of, their knowledge of the contest, and decide only whether or not to the incur the cost (either effort or monetary) of submission, as in arts or literature contests\footnote{As one example, a number of travel photography contests are frequently hosted by both government agencies and private enterprises such as National Geographic---participants in these contests likely do not undertake their travel to produce a photo for such a contest, but rather choose whether to submit an eligible photo or not.}. 
\item Second, there are environments where producing an entry indeed incurs a cost, but this cost essentially does not vary with the quality of the entry---for example, contributing an answer on expertise-based\footnote{(\ie, where the quality or value of an answer depends much more on the expertise of the answerer than her effort, such as health or law forums)} online Q\&A forums incurs a cost\footnote{(the cost of logging in and posting the answer on the forum)} which is essentially independent of the contribution's quality. 
\item Finally, there are `best-effort' environments where producing an entry does incur a cost, and incurring a higher effort cost does increase the quality of the resulting submission, but where contestants do not `game' \ntb{their} effort---having made the decision to participate, a contestant will work to the best of her ability to produce the best entry she possibly can. Any contest-like scenario where entry is driven by intrinsic motivation for the task at hand can fall into this category. This category includes offline contests for, say, a performing art. It also includes online contests like Quirky, as well as citizen-science and user-contributed content driven communities.
% The grammatical structure of the last sentence was unclear to me, so I couldn't figure out how to reword.
Indeed, there is a growing literature suggesting that users in several online contribution settings do not strategize on effort on a task in response to incentives, although they might choose whether or not to undertake the task based on the promised rewards~\cite{KRbook}\footnote{\ntb{\citet{KRbook} summarize} this as a `design claim': ``With task-contingent rewards for small, discrete tasks, larger rewards will motivate people to take on tasks, but will not motivate higher effort on accepted tasks.''}; for instance, \cite{JKC10} find in a field experiment on Google Answers that participation, but not quality, is sensitive to the prize offered to winning answers.
 \EIT  
There are clearly a vast range of real-life scenarios, therefore, where potential contestants do strategize about whether or not to participate in \ntb{a} contest, but not about the quality of their submissions. How do incentives and equilibrium outcomes change when agents only make strategic participation choices, rather than strategic effort choices, to maximize their expected utilities\ntb{? Specifically,} what allocation of prizes for each rank leads to the 
`best' elicited set of submissions in equilibrium? \\
%induces equilibrium participation profiles leading to the `best' elicited set of submissions? \\

\subsection{Our contributions} 
We explore the idea of `behavioral' mechanism design via the problem of designing optimal contests for {\em simple} agents, where a principal seeks to allocate a total prize to optimize the qualities of elicited submissions when potential contestants only strategize about whether to participate in the contest. 
%but not over the quality of their submissions. 
Specifically, consider a population of $n$ potential contestants with types $(q_i,c_i)$ drawn from a known joint distribution $F(q,c)$, where an agent with type $(q_i,c_i)$ produces output of quality $q_i$ at a cost $c_i$ if she participates in the contest. Suppose there is a principal who can distribute a total prize $V$ amongst agents based on their relative ranks; here $V$ might be either a monetary prize budget, or a non-monetary resource such as display space on a webpage that translates to attention rewards. Each agent decides whether or not to participate by comparing her expected prize from participating (which depends on the rank of her submission's quality $q_i$ relative to that of the other agents who also choose to enter) against her cost of participation $c_i$.  What allocation of $V$ amongst the $n$ ranks induces equilibrium participation decisions that lead to the best outcomes? 

Participation-only strategic choices can be thought of as a special case of a strategic effort choice model, where agents are restricted to a binary choice between the maximum possible effort and zero effort. While it might seem, at first glance, that reducing agents' choices from an interval to the endpoints of that interval ought not change the nature of the optimal contest design problem much, or at least only make it simpler, this turns out to be far from the truth---agents' incentives are significantly altered by the fact that any competitor who does choose to participate will produce her `best possible output' instead of `adjusting' her effort  (and correspondingly the submission quality that all other participants must compete against) to some suitable point in the interior of the interval. 
We note here an interesting conceptual parallel with discrete versus continuous optimization: moving from continuous to discrete optimization problems where variables can only take binary (or integer) values fundamentally changes the nature of the algorithm design question, leading to an entirely different body of techniques for algorithm design and, in many cases, to computationally intractable problems. A second factor that contributes to the complexity of the strategic problem in these settings is that agents can, in general, have two-dimensional types: as discussed below, this significantly complicates the nature of equilibria, necessitating new tools to bound the outcomes elicited in an equilibrium of a contest.

We first study the optimal contest design problem in settings with homogenous participation costs $c$, so that agents' types \ntb{$(q_i,c)$} are essentially one-dimensional: here, the contest that elicits the highest equilibrium participation is optimal for any objective that is an increasing set function of the elicited submission qualities.  We show that this optimal contest is always a {\em simple contest} which awards equal prizes of $V/j^*$ each to the top $j^*$ contestants, where the optimal number of prizes $j^*$ scales with $\vc = V/c$ as $\vc - \Theta(\sqrt{\vc/\log(\vc)})$. This contrasts with the optimal contest structure in comparable models with strategic effort choices, where the optimal contest is either a winner-take-all contest or may award unequal prizes, depending on the curvature of agents' effort cost functions~\cite{MS01}.  This means, for example, that in online user-generated content environments where users do decide whether or not to contribute, but do not strategize over the qualities of the content they produce, handing out an appropriately chosen number of identical badges might lead to better outcomes than reward structures---such as ranked top-contributor lists---inducing unequal social-psychological rewards. As another example, a suitably chosen number of equal `merit' prizes in a travel photography contest might elicit superior submissions than a structure that awards unequal first, second, and third prizes\footnote{For example, the National Geographic 2014 contest awards 3 unequal top prizes and 7 identical merit prizes: http://travel.nationalgeographic.com/travel/traveler-magazine/photo-contest/2014/.}.

We next address the general case when agents have heterogenous costs $c_i$ that can be arbitrarily correlated with the qualities $q_i$ of their submissions. Here, agents' types $(q_i,c_i)$ are inherently two-dimensional, unlike in the homogenous cost model in \S \ref{s-homo}, or in strategic effort models in the contest design literature  (where agents' types are parametrized by a one-dimensional ability that either scales agents' cost functions or their outputs per unit effort). With two-dimensional types, much of the structure that is central to the equilibrium analysis of contests with one-dimensional agent types---where equilibrium action choices are typically monotone in this type---vanishes, making it difficult to explicitly characterize equilibria, and therefore to identify the contest structure that yields the `best' equilibrium outcome. 

Since the design of the optimal contest can, in general, vary depending on the choice of objective function in this setting, we focus on contests that maximize the quality of the {\em best} elicited submission in equilibrium. 
Our main result in this model concerns the winner-take-all
contest, which awards the entire prize $V$ to the top entry. 
A winner-take-all contest does not necessarily optimize the expected
quality of the best elicited submission, as shown,
for example, by our analysis of the case of homogeneous participation
costs: we show, however, that the
winner-take-all contest always achieves a bounded approximation 
factor~\footnote{One could question whether such approximation results
are meaningful, since it is unclear what it even means to quantify
the quality of elicited submissions numerically. 
Our approximation result is at least partly immune from this
critique for the following reason. First note that the
critique applies only to the way we evaluate the quality of a 
mechanism's outcome in equilibrium, not to the way we model
mechanisms, agents, and equilibria. This is because 
the mechanisms we consider 
are rank-order contests which make only ordinal comparisons of 
quality, so the behavior of a mechanism and its
set of equilibria are unaffected by monotonic reparameterizations of the 
quality scale. For essentially the same reason, our approximation
result about the winner-take-all contest remains valid under any
monotonic reparameterization of quality; the reparameterization
does not affect the set of equilibria of the contest, so the 
conclusion of any theorem about that set of equilibria cannot 
be sensitive to whether we have reparameterized the quality scale.}
with respect to the optimal contest. More precisely,
% Our main result in this model shows that the winner-take-all contest
% is a 3-approximation of the optimal contest---
% that is, 
the expected value of the highest-quality entry elicited in any pure-strategy equilibrium of the winner-take-all contest is no smaller than one-third of the expected maximum quality elicited in the best equilibrium of an optimal contest $\M^*$. 
One interpretation of this result is that the designer's
problem 
% of optimizing the quality of the best elicited submission 
has an approximately detail-free solution:
without knowing the joint distribution of $(q_i,c_i)$,
the designer can run a winner-take-all contest and
thereby guarantee, in equilibrium, an outcome not
much worse than the best equilibrium achievable by any contest.
The proof of this result hinges around a `sub-equilibrium' lemma, which establishes a stochastic dominance relation between the distribution of qualities elicited in an equilibrium and in a {\em sub-equilibrium}---a strategy profile that is a best response for all agents who choose to {\em participate}, but not necessarily a best response for non-participants---of the winner-take-all contest, and may have broader applications to characterizing equilibrium performance in settings where it is difficult to explicitly solve for equilibria. 

\subsection{Related work}
\label{s-relwork}
There is now a tremendously large literature on the economics of contests. The seminal work of~\citet{Lazear81} suggesting that contests, or rank-order tournaments, can serve as an efficient scheme to \ntb{incentivize} effort from strategic agents has given rise to a vast literature on contests as a means for effort elicitation, see e.g.~\cite{Green83,Che03}. 
The problem of optimal contest design---how to allocate rewards to elicit the most desirable outcomes, posed by Sir Francis Galton back in 1902 and formally addressed as early as~\cite{Glazer88}---has been studied in a vast variety of settings, encompassing heterogeneous (e.g.~\cite{MS01, MS06, CHS11}) and homogenous (e.g.~\cite{Taylor95, GM12}) agent populations, risk-neutral and risk-averse preferences over prizes, non-monetary rewards~\cite{MSS07,GM12}, as well as various models of information and observability of output, and various objectives and constraints faced by the principal running the contest. 

While very large, this literature---to the best of our knowledge---studies contests where agents make strategic effort choices that (deterministically or stochastically) affect the quality of their submissions, and does not address the problem of optimally designing contests for agents who make only strategic participation choices, a difference that significantly alters the nature of agents' incentives as discussed in \S \ref{s-intro}. The model closest to ours from this literature is perhaps that of ~\citet{Taylor95}, where agents can make a random draw of output quality from a distribution $F$ at a cost of $c$, similar to our model in \S \ref{s-homo}. However, agents repeatedly make this choice over $T$ periods (the duration of the contest), so that the central incentive problem faced by an agent  in~\cite{Taylor95} is whether to make an {\em additional} draw at this cost. This repetition of the draw essentially mimicks a strategic choice of effort, albeit under different informational circumstances than in most other economic models of contests. 

There is also a growing literature on modeling and analyzing incentives in online competitive environments such as crowdsourcing contests and online user-contributed content (see, e.g.,~\cite{DV09,AS09,CHS11,GH12, GM12,CJ12, EG13,ISS13}). As in the contest design literature, almost all of this work models agents as making non-binary strategic effort (or quality) choices, and asks how to design incentives to elicit desired effort profiles from contestants. We note here  that~\citet{CJ12} find---albeit in a completely different model from ours---that for specific distributions of agent abilities that satisfy a certain condition, the {\em efficient}, \ie, first-best outcome (accounting for both the principal's value and agents' costs to effort), would be for some number of agents to exert maximum effort and for the remaining agents to exert no effort;~\citet{CJ12} then address the question of implementing the efficient outcome in equilibrium, \ie, when agents strategically choose effort in response to the contest design, and prove both existence and impossibility results. While these `all or nothing' effort choices bear an interesting resemblance to our model with agents who strategize only about participation, this resemblance is only superficial: the all-or-nothing choice in our model is an {\em assumption} about agents' strategic behavior, while it arises as a structural property of efficient outcomes under certain  conditions in~\cite{CJ12}.
\iffalse 
{We note here  that~\cite{CJ12} find---albeit in a completely different model from ours---that for specific distributions of agent abilities that satisfy a certain condition, the {\em efficient}, \ie, first-best outcome (accounting for both the principal's value and agents' costs to effort), would be for some number of agents to exert maximum effort and for the remaining agents to exert no effort, which bears an interesting, although very superficial and restricted, resemblance to `participation-only'  outcomes.~\cite{CJ12} then address the question of implementing the efficient outcome in equilibrium, \ie, when agents strategically choose effort in response to the contest design, and prove both existence and impossibility results.}  
\fi 

The most closely related work to ours from this literature is~\cite{GH12} which investigates the implementability of optimal outcomes in widely used `best-contribution' contests, in a model where agents with randomly drawn heterogenous abilities make participation-only choices incurring a uniform cost. While that model corresponds precisely to the model in \S \ref{s-homo} of this paper, the nature of the questions addressed are fundamentally different---\citet{GH12} {\em analyze} the implementability of optimal outcomes in a specific contest structure {\em without} resource constraints; in contrast, we ask how to {\em design} the optimal contest under a given resource constraint, and identify which amongst several possible contest structures achieves the best outcome subject to this resource constraint.  

We note that the model of simple agents who strategically reason only about participation choices is not new to this paper---in addition to~\cite{GH12}, it appears at least in~\cite{GL13} in the context of privacy-sensitive users who make simple participation-only decisions about whether to share their data (although again in the context of equilibrium analysis rather than of optimal design), as well as possibly in several other settings in the vast economics literature. In the same spirit, we note that the general idea of moving from ideal to more realistic models is also not new to this paper: in a broad sense, our work belongs to a line of previous literature---albeit in completely different settings, with very different questions and models---that is driven by differences between the idealized models in traditional economic analyses and `real' environments. This includes, for instance, the work on behavioral game theory~\cite{Camerer03}, and building more realistic game-theoretic models of agent behavior as in~\cite{WLB14}, as well as designing mechanisms for more realistic assumptions about agents' information or preferences, such as the literature on deliberative agents initiated in~\cite{LS05}.

\section{Model}
\label{s-model}
%% \textcolor{red}{Ability versus output: notation, terminology. \\
%% p versus q = 1-p. Clumsiness earlier versus later.\\
%% Indexing: Agents by $i$, prizes by $j$. $k$ for summation indices.}\\

We consider a model with $n$ potential participants, each of whom strategically chooses only whether or not to participate in a given rank-order contest $\M$. An agent who decides to participate will produce an output of quality $q_i$, incurring a cost $c_i$, where agents' quality-cost pairs are drawn from a known joint distribution $F$; an agent makes her strategic participation choice based on whether her expected benefit from participation is large enough to cover her cost. Naturally, each agent's choice depends on the participation choices of the remaining agents, since these determine the relative rank of her output $q_i$ and therefore her prize. We will be interested in designing contests $\M$ that maximize some increasing function of the qualities of submissions from the agents who choose to participate in the contest {\em in equilibrium}. We describe the model formally below. \\
  
\noindent{\bf Agents.} There is a pool of $n$ agents who are all potential participants in a contest. Agents are {\em simple}---they only strategically choose whether or not to {\em participate} in the contest, but do not strategically adjust their efforts, or equivalently the quality of their outputs, in response to the contest structure and environment. That is, {\em if} an agent decides to participate, she puts in a predetermined amount of effort incurring a corresponding cost, leading to some corresponding output quality; if her effort is too costly relative to her expected benefit from the resulting quality of output, she will simply not participate. 

We model the fact that agents might have different abilities or skills relevant to the task posed by the contest via the quality of their output, as well as the cost they incur to produce this output---we denote the quality of the output that agent $i$ can produce, if she chooses to participate, by $q_i$, and her associated cost by $c_i$. (An agent who does not participate produces no output, and incurs no cost.) An agent $i$ is fully described by her quality-cost pair $(q_i, c_i)$; we assume that the tuples $(q_i,c_i)$ \ntb{for each of the $n$ agents are i.i.d.\ samples} from a joint distribution $F(q,c)$ which is common knowledge to all agents and the contest designer. \ntb{We additionally assume that the marginal
distribution of qualities has no mass points: for every $q_0$, the event $q=q_0$
has zero probability under $F$.}\\
% \tr{Smoothness assumptions on $F$?}
% \tr{Equating ability and output.} \\

\noindent {\bf Contests.} We suppose that the principal can observe the {\em rank-ordering} of the qualities $q_i$ of participants' submissions, and correspondingly consider rank-order mechanisms that award prizes to agents based (only) on the rank of their outputs. We use $\M\Mv$ to denote the rank-order mechanism that awards a prize $v_j$ to the agent with the $j^{\mathrm{th}}$ rank, \ntb{when submissions are arranged in order of decreasing quality with ties broken at random}\footnote{Note that our assumption that the marginal distribution of $q$ has no point masses implies that almost surely, the tie-breaking rule need not be invoked.}. A mechanism $\M\Mv$ is monotone if higher-ranked agents receive larger prizes, \ie, if 
$$v_1 \geq v_2 \geq \cdots \geq v_n,$$ 
and monotone nonnegative if, in addition, $v_n \geq 0$. We will use the term contest to mean a monotone nonnegative rank-order mechanism henceforth in the paper.\\ 

\noindent {\bf Utility, strategies, and equilibrium.} Agents are risk-neutral with quasi-linear utilities: an agent's utility, if she chooses to participate in a contest $\M\Mv$, is the difference between her expected prize (determined by her rank, which depends on how her quality $q_i$ ranks relative to the random draws $q_{i'}$ of the remaining agents who choose to participate) and her cost $c_i$ to produce the output $q_i$. An agent who does not participate receives no prize and incurs no cost, and so obtains utility $0$. 

A strategy prescribes whether or not an agent with a given quality-cost draw should participate in a contest. Formally, a (pure) {\em strategy} $\strat$ is a function mapping a quality-cost tuple $(q,c)$ to a binary participation decision specifying whether or not to participate in a mechanism $\M$; a  player $(q_i,c_i)$ following strategy $\strat$ participates in $\M$ if and only if $\strat(q_i,c_i)=1$. A pure strategy profile $(\strat_1,\ldots,\strat_n)$ is a vector of strategies, one for each of the $n$ players; a {\em symmetric} pure strategy profile is one where $\strat_i = \strat$ for all $i \in [n]$ and is denoted  $\vec{\strat}$. Finally, a pure strategy profile in which $\strat_{i'}=\strat$ for all $i' \neq i$, but 
$\strat_i$ may differ from $\strat$, will be denoted by $(\strat_i,\vec{\strat}_{-i})$.

Agents make their strategic participation choices based on the mechanism $\M\Mv$ announced by the principal, the strategies $\vec{\strat}_{-i}$ of other agents, and their knowledge of the distribution $F$ of quality-cost tuples $(q,c)$ and the number of agents $n$. An agent $i$ with quality and cost $(q_i,c_i)$ decides whether to participate or not by comparing her expected utility from producing output $q_i$ at cost $c_i$ against the utility of $0$ from non-participation; we assume that an agent who is indifferent between participation and non-participation always breaks ties in favor of participation\footnote{\ntb{In this paper, we always adopt assumptions on the joint distribution of $(q_i,c_i)$ (such as no point masses) that are sufficient to ensure that agents only employ this tie-breaking rule with probability $0$, so that this assumption does not affect our qualitative conclusions, but only makes our results easier to state.}}.

A symmetric pure-strategy profile $\vec\strat$ constitutes a Bayes-Nash equilibrium if no agent can profitably deviate from the recommendation of the strategy $\strat$ given her draw $(q_i,c_i)$; that is, (i) an agent with $\strat(q_i,c_i) = 1$, \ie, who participates when playing according to $\strat$, obtains non-negative expected utility assuming that the remaining $n-1$ agents with $(q,c)$ drawn from $F$ also play according to $\strat$, and (ii) the expected utility of an agent with $\strat(q_i,c_i) = 0$, if she were to participate while the remaining $n-1$ agents play according to $\strat$, is negative. That is, no participant can beneficially deviate by not participating, and no non-participant would wish to participate.

For a given symmetric strategy profile $\vec \strat$, the set of participants in $\M$ is a random variable constructed by choosing $n$ i.i.d.\ draws of $(q_i,c_i)$ from $F$, and then applying $\strat$ to each of these $(q_i,c_i)$ tuples.
We use $\spart(\strat)$ to denote the random set of participants when all $n-1$ agents follow the strategy profile $\strat$. Note that $\spart$ depends only on the strategy profile chosen by agents, and does not depend at all on the mechanism $\M$.  \\

\noindent {\bf Objective and constraints.} We will be interested in the design of optimal contests, \ie, an allocation of rewards $\Mv$ that maximizes some function of the qualities of submissions from the agents who choose to participate in a symmetric equilibrium\footnote{All contests that we consider in this paper have a symmetric equilibrium which is unique, and we restrict our attention to symmetric equilibria.} of the contest, subject to some constraint on the available rewards $v_j$. (Note that these rewards $v_j$ may be either  monetary, or non-monetary social-psychological rewards; see \S \ref{s-intro}.)  While there are \ntb{various} possible constraints on the rewards\footnote{For instance, if the $v_i$ are attention rewards associated with content being displayed down a webpage, a natural constraint is a maximum per-slot attention constraint of the form $v_j \leq V_j$~\cite{GM12}.}, we will focus in this paper on a 
sum constraint: 
$
  \sum_{j = 1}^n v_i \leq V.
$  
We assume throughout that the rewards, or prizes, $v_j$ for each rank $j$ are announced {\em prior} to the contest. Note this means that there might be leftover prize money for any particular instantiation of the agents' $(q_i,c_i)$ pairs and strategic choices: depending on how many agents actually participate,  the total sum of rewards given out to participating agents might be strictly less than $V$. That is, we consider rank-order mechanisms where the principal commits to a single allocation of $V$ amongst the $n$ ranks a priori, and cannot redistribute $V$ (or equivalently, any leftover prize money) depending on the participation actually elicited, in contrast to hypothetical mechanisms where the principal  might announce a full menu of prizes $((v_1^1),  (v^2_1, v^2_2), \ldots, (v^n_1, \ldots, v^n_n))$ for each possible level of participation $k = 1, \ldots, n$. 

The outcome of a contest is determined by the output qualities $q_i$ of agents who choose to participate in the contest. Let $\qpart(\strat) \in R^{n}$ denote the (random) vector consisting of the  ordered output qualities of agents who participate when all agents play according to the strategy profile $\strat$.  Formally, define an agent $i$'s output under $\strat$ to be $q_i$ if $i$ participates (\ie, $\strat(q_i,c_i) = 1$) and $0$ if she does not ($\strat(q_i,c_i) = 0$); the (random) vector $\qpart(\strat) $ then holds the $k^{\mathrm{th}}$ largest of these $n$ outputs $q_i  \cdot  \strat(q_i,c_i)$ in its $k^{\mathrm{th}}$ component. 

The principal seeks to design mechanisms that maximize the expected value of some objective function $f$ of the output qualities elicited in equilibrium.
% {I chose not to reiterate `symmetric in this sentence, because we're stating it in the following one.} 
We define an optimal contest for a given objective $f$ to be any mechanism $\M\Mv$ such that $\M$ has a symmetric pure-strategy equilibrium that maximizes $E[f(\qpart(\strat^*(\M)))]$\footnote{As before, the expectation is over random draws of $(q_i, c_i)$ from $F$.} over all symmetric pure-strategy equilibria $\strat^*(\M)$ of all feasible contests $\M$. Note that it is not clear, {\em a priori}, that there exists any mechanism with a symmetric pure-strategy equilibrium, nor that a given mechanism necessarily has such an equilibrium or that the equilibrium is unique when it exists. In Sections~\ref{s-homo} and~\ref{s-hetero}, we will prove existence and uniqueness of symmetric equilibria under the mild hypotheses that the distribution of $q_i$ (in \S \ref{s-homo}) and the joint distribution of $(q_i,c_i)$ (in \S \ref{s-hetero}) are absolutely continuous with respect to Lebesgue measure, \ie, they assign probability zero to every measure-zero set. 
% (Note that whether a mechanism $\M$ is optimal or not depends on it's `best', rather than `worst', equilibrium--- we use this definition since where we derive the optimal contest (\S \ref{s-homo}), all contests have a unique pure-strategy equilibrium, whereas we approximate the optimal contest in \S \ref{s-hetero} where using the best equilibrium to define optimal contests leads to a stronger result.) \ntb{Edit this whole paragraph once the equilibrium existence thing is straightened out.}

Throughout, we will be interested in objective functions that are {\em increasing} functions of the qualities of the elicited outputs: a function $f$ is {\em increasing} if $f(\qpart_1) \geq f(\qpart_2)$ whenever the vector $\qpart_1 \succeq \qpart_2$. Informally, a function $f$ of the set of elicited output qualities is increasing if higher-quality outputs are (weakly) preferred to lower-quality ones, and increasing participation to a superset of existing participants also makes the principal happier. 
We note that this assumption---that `more is better', or equivalently, free disposal of output---is not entirely without loss of generality: it does not capture, for instance, search costs associated with sorting or finding better outputs in a larger set of outputs, or equivalently says that such costs are zero or negligible compared to the benefit of having a larger set of outputs. Nonetheless,  the assumption of an increasing objective is reasonable in a vast number of settings, and essentially all the objective functions used in the (now vast) literature on contest design, including the two most commonly used objectives, namely  the maximum output $\max_{i \in \spart} q_i$ and the sum of outputs $\sum_{i \in \spart} q_i$, are such increasing functions.

\section{Homogenous participation costs} 
\label{s-homo}
We begin by considering a special case of the distribution $F(q,c)$ where all agents have homogenous costs of participation, \ie, $c_i = c$ for all $i$. A number of competitive environments can be reasonably described by a model with such homogenous participation costs---any contest where the participation cost $c_i$ is primarily the cost of submission (either monetary, such as an entry fee, or the effort cost) rather than production (\S \ref{s-intro}), or where the cost of producing an entry is largely independent of its quality (as in online content contribution settings where users differ in their abilities to contribute but incur similar contribution costs irrespective of quality, or in time-bound contests where the perceived cost is the time cost of participating in the contest), are well captured by a model with homogenous costs $c_i = c$.

We first  analyze symmetric equilibria\footnote{Since we restrict attention to symmetric equilibria throughout this paper, we henceforth omit the modifier ``symmetric'' when discussing equilibria.} of contests when agents have homogenous costs in \S \ref{s-homo-eq}, and show that the equilibrium in any contest always consists of a threshold strategy. This means that the optimal contest for any increasing objective function of the designer is identical, and corresponds to a contest with the lowest threshold. We use this equilibrium analysis and an LP formulation of the contest design problem to prove our main result (\S \ref{s-simple}), which is that optimal contests are {\em simple}: an optimal contest awards an equal prize to some number $j^*$ of the highest-ranked contestants, and no prize to the rest. (Note that this structure is in contrast with the optimal contest structure in comparable models with strategic {\em effort} choices, where the optimal contest is either a winner-take-all contest or may award unequal prizes, depending on the curvature of agents' effort cost functions~\cite{MS01}.) 
The implications of this result for contest structures in various practical competitive environments are discussed in \S \ref{s-intro}\footnote{See ``Our contributions''.}. 
%% The fact that the optimal contests for simple agents with homogenous costs are simple implies, for example, that in online user-generated content environments where users do decide whether or not to contribute, but do not strategize over the qualities of the content they produce, handing out an appropriately chosen number of identical badges might lead to better outcomes than reward structures---such as ranked top-contributor lists---inducing unequal social-psychological rewards\footnote{As another example, a suitably chosen number of equal `merit' prizes in a travel photography contest might elicit superior submissions than a structure that awards unequal first, second, and third prizes. Interestingly, we note that the National Geographic 2014 contest awards 3 unequal top prizes and 7 identical merit prizes: http://travel.nationalgeographic.com/travel/traveler-magazine/photo-contest/2014/.}. 
Finally, we investigate how the optimal number of prizes $j^*$, as well as the expected participation, vary with the `scale' $V/c$ of the contest in \S \ref{sec:compstat}. 
% Due to page limits, most proofs have been
% omitted from this version of the paper and will be included in a full
% version posted on the authors' websites.} \bkmargincomment{Would be convenient
% to provide a citation with a link to someplace on the web where people can
% download the full version.}

Throughout this section, we will use $F(q)$ to denote the marginal distribution from which agents' qualities $q_i$ are drawn. We will also assume that $\frac{V}{n} < c$, \ie, there is not enough prize money to incentivize all agents to participate, since otherwise the optimal contest design problem is trivial: the contest with $v_j = \frac{V}{n} \geq c$ achieves full participation, since all agents receive a prize that covers or exceeds their cost to participation.
 
\subsection{Equilibrium analysis} 
\label{s-homo-eq}
We first analyze equilibrium behavior in a general contest $\M\Mv$ with simple agents who have homogenous costs. We show that with homogenous costs, the equilibria for any contest are threshold-strategy equilibria, where all agents with qualities $q_i$ above some threshold participate, and the remaining agents do not. Before proving the result, we first introduce the function $c_{\M}(p)$ defined by
\BEQ \label{e-cm}
c_{\M}(p) = \sum_{j=1}^n v_j {n-1 \choose j-1} p^{j-1}(1-p)^{n-j}. 
\EEQ 
This function will turn out to be central to our analysis.
% , and relates to 
Its value denotes the expected prize earned by a participant in the contest $\M\Mv$ when each of the other $n-1$ agents has probability $p$ of participating and placing ahead of that participant. Of course, the value of $p$ depends on the type of that agent---her output quality $q_i$---as well as the strategies upon which each of the remaining $n-1$ contestants base their participation decisions, so that $p = p(q_i, \strat_{-i})$. When all agents play according to a common {\em threshold} strategy, though, if an agent with quality $q_i$ participates then this probability of another agent participating and beating her is precisely $p = 1-F(q_i)$, since $i$ participates only if her only quality meets or exceeds the threshold.  
%\tr{This function represents the expected prize earned by a participant
%in the contest $\M\Mv$, when each of the other $n-1$ agents has
%probability $p$ of placing ahead of that participant.}\tr{(AG: A little misleading: Note %that p depends on agent $i$, as well as other agents' strategies.)}

The following lemma will be crucial to our analysis. 
\begin{lemma} 
\label{l-qc1to1}
For any monotone rank-order mechanism $\M = \M(v_1,\ldots,v_n)$, 
$c_{\M}(p)$ is non-increasing in $p$. 
If $v_1 \ne v_n$, then $c_{\M}(p)$ is strictly decreasing in $p$.
\end{lemma} 
\proof
%% A sketch of the proof follows, and a more rigorous proof
%% is provided in % Appendix~\ref{s-homo-eq-proofs}.
%% the full version of this paper.
As stated above,
% It is evident from~\eqref{e-cm} that 
$c_{\M}(p)$ is equal to the
expected prize gained by a contestant in contest $\M$ with $n-1$ other agents, when each of these agents (independently) places ahead of her with probability $p$. Intuitively, as we increase $p$, this contestant's (random) rank for larger values of $p$ stochastically dominates the rank for smaller values of $p$, and the lemma follows since 
prize is a decreasing function of rank. 

More rigorously,
for $p_0 < p_1$, consider an urn containing three types of 
balls: red balls with probability $p_0$, white balls with
probability $p_1-p_0$, and blue balls with probability $1-p_1$.
Consider a random experiment in which $n-1$ balls are sampled
independently (with replacement) from the urn, and $j_0$ is 
defined to be the number of red balls sampled while
$j_1$ is defined to be the number of sampled balls that
are either red or white. From the formula for $c_{\M}(p)$
one can see that $c_{\M}(p_0)$ is equal to the expected 
value of $v_{j_0+1}$ while $c_{\M}(p_1)$ is equal to the 
expected value of $v_{j_1+1}$. Since $j_0 \leq j_1$ pointwise
and $v_1,\ldots,v_n$ is a non-increasing sequence,
it follows that $c_{\M}(p_0) \leq c_{\M}(p_1)$ as claimed.
There is a positive probability that $n-1$ white balls are
drawn from the urn, and in the event that this happens
$j_0+1=1$ while $j_1+1=n$. Thus, assuming $v_1 > v_n$
we have $c_{\M}(p_0) < c_{\M}(p_1)$ which confirms
the second claim in the lemma.
\endproof

% Now we prove 
The next proposition asserts that rank-order mechanisms have unique 
equilibria in threshold strategies. The proof is deferred to
Appendix~\ref{s-homo-proofs}.
%The proof is quite simple,
%and is given in the full version of the paper.
\begin{proposition} \label{p-eq-homo}
For any monotone rank-order mechanism $\M$, there is a symmetric equilibrium in threshold strategies where every agent with quality $q_i \geq \at$ participates, and every agent with quality $q_i < \at$ does not participate; the equilibrium threshold $\at$ satisfies $c_{\M}(1-F(\at)) = c$.
This is the unique symmetric equilibrium of $\M$.
\end{proposition}

\subsection{Optimal contest design: LP formulation} 
\label{s-lp}
The fact that there is a unique equilibrium in threshold strategies for every monotone rank-order mechanism $\M = \M\Mv$ means that the (equilibrium) vector of outputs $\qpart_1$ in a mechanism $\M_1$ with  threshold $\at_1$ dominates the vector of outputs $\qpart_2$ in a mechanism $\M_2$ with threshold $\at_2 > \at_1$, for every possible realization of the qualities $q_i$: every agent who participates in equilibrium in the second mechanism also participates in the first, and there could possibly be additional agents who participate in $\M_1$ but not in $\M_2$. This means that the problem of choosing the optimal mechanism for any objective $f(\qpart)$ which is increasing in the vector of outputs $\qpart$ 
% (Definition \ref{d-inc}) 
becomes very simple: we want to find the mechanism with the allocation of rewards $\Mv$ that leads to the lowest equilibrium threshold $\at(\M)$.  
%Since {\em all} agents with qualities greater than the threshold participate, 
%\label{p-eq-homo}

Consider any mechanism $\M\Mv$, and let $p = 1-F(\at)$ where $\at$ is the equilibrium threshold in $\M\Mv$ (Proposition~\ref{p-eq-homo}). Since all agents with qualities $q_i \geq \at$ participate in $\M$ in equilibrium, we can interpret $p$ as the ex-ante probability (\ie, prior to observing the draw of $q_i$) with which each agent will participate in the mechanism. We want to distribute the total prize $V$ into $\Mv$ so as to maximize this participation probability $p = 1-F(\at)$, since maximizing this participation probability maximizes any increasing objective $f$ of the principal running the contest. \\

\noindent{\bf Feasibility.} First, note that the problem of {\em deciding} whether or not a {\em given} participation rate $p$ is feasible for a given total prize $V$, and participation cost $c$, can be written as a feasibility LP in the variables $\Mv$: 
\BEQ
\begin{array}{rl}
\label{e-lp-f}
%\mbox{minimize} && \sum_{i \in N}   p_i \sum_{r \in N} x_{ir} \cdot l(i,r) \\ 
v_1 \;\geq\; v_2 \;\geq \ldots \geq\; v_n & \geq \; 0, \\
\sum_{j=1}^n v_j & \leq\; V,\\
\sum_{j=1}^n v_j {n-1 \choose j-1} p^{j-1} (1-p)^{n-j} & =\; c. % \label{e-lp-inc}
\end{array}
\EEQ
The first two constraints say that we restrict ourselves to monotone nonnegative mechanisms $\M$ that distribute a total prize of at most $V$. The third constraint is the {\em incentive} constraint; it accounts for the fact that agents make strategic (albeit simple) choices in response to the incentives created by the mechanism $\M\Mv$, and ensures that $p = 1-F(\at)$ actually describes the equilibrium threshold in $\M$. 

If this LP is feasible, there exists an allocation of the total reward $V$ into rewards $\Mv$ that elicit an equilibrium participation probability of $p$; if not, no contest with total available prize $V$ can achieve a participation rate of $p$ when agents have cost $c$ to participation. \\

\noindent{\bf Optimality.}  The feasibility LP asks {\em whether} it is possible to achieve some level of participation $p$ given a total reward $V$, when agents have participation costs $c$.  The question we would actually like to ask is {\em what} is the highest participation rate we can achieve given the pair $(V,c)$, and how---\ie, using what allocation scheme $\Mv$? Since the incentive constraint---the third constraint in~\eqref{e-lp-f}---is nonlinear in $p$, we cannot write the problem of maximizing $p$ over all allowable choices of $\Mv$ for given values of $c$ and $V$ as an LP. However, the monotonicity between $c$ and $p$ established in Lemma~\ref{l-qc1to1} will allow us to approach this question via an LP anyway. 

%We now address this question.  
%To answer this question, we need a lemma that relates $c$ to $q$. 
 
Fix a mechanism $\M = \M\Mv$: for every $p$, the function $c_{\M}(p)$ defined in~\eqref{e-cm} gives the precise 
participation cost at which $\M\Mv$ has equilibrium participation rate $p$. 
%Lemma~\ref{l-qc1to1}, which asserts the monotonicity of $c_{\M}(p)$, implies that for any fixed rank-order mechanism $\M\Mv$ with $v_1 \neq v_n$, in order to achieve a strictly higher participation probability $p$, the cost of participation $c$ must strictly decrease. 
%AG: It's really not required to mention this here, for this part of the reasoning. 
Now for a {\em given} $p$, consider the following LP in the variables $\Mv$: %; note that the objective is precisely $c_{\M}(q)$. 
\BEQ
\begin{array}{cl}
\label{e-lp-o}
\mbox{maximize}  &\sum_{j=1}^n v_j {n-1 \choose j-1} p^{j-1} (1-p)^{n-j}\\ 
\mbox{s.t.} & v_1 \geq v_2 \geq \ldots \geq v_n \geq 0, \\
&\sum_{j=1}^n v_j = V. 
\end{array}
\EEQ
Note that the objective of this LP is precisely $c_{\M}(p)$. Therefore, for a given $p \in [0,1]$, the solution to this optimization problem gives the largest participation cost $c$ at which a participation rate of $p$ can be supported in equilibrium by optimizing over how the prize $V$ is distributed across ranks. 

Lemma \ref{l-qc1to1}, which asserts the monotonicity of $c_{\M}(p)$, allows us to use this LP to address the question of {\em optimal} contests, \ie, contests that achieve the best possible participation rate $p$ 
%(and therefore the expected value of the outcome, for any increasing objective $f$) 
for a given total prize $V$ and participation cost $c$, as follows. 
Define the function $c^*(p)$ to be the optimal value of the LP (\ref{e-lp-o}) for each $p \in [0,1]$: $c^*(p) = \max_{\M\Mv} c_{\M}(p)$ describes the maximum participation cost $c$ at which a participation rate of $p$ can still be supported in equilibrium by appropriately choosing the most favorable reward allocation $\Mv$. 
From Lemma \ref{l-qc1to1}, the function $c^*(p)$ is strictly decreasing, since it is the pointwise maximum of strictly decreasing functions. 
(Our assumption that $\frac{V}{n}< c$ implies that $v_1 \neq v_n$ for any contest that elicits non-zero participation, since if $v_1 = v_n$, the prize $v_j \leq \frac{V}{n}$ for all ranks $j$ so that no agent participates).
Hence $c^*(p)$ is also invertible; define the inverse of this function to be $p^*(c)$. The function $p^*(c)$ is the highest participation rate  that can be supported in equilibrium with a total prize $V$ when agents have participation cost $c$, and this rate is achieved by any contest that maximizes the objective in (\ref{e-lp-o}) for $p = p^*(c)$. 

Note that the invertibility of $c^*(p)$ is important {\em not} because we cannot answer the question of the best achievable $p^*$ for a given $(V,c)$ without it, but rather because it allows us to say that an allocation $\Mv$ achieving the optimum of (\ref{e-lp-o}) for any specified $p$ is indeed an optimal contest (for the value of $c$ corresponding to the LP optimum). Without invertibility, we could still define $p^*(c)$ as the largest $p$ for which $c^*(p)$ was equal to $c$, to answer the question of the best achievable $p$ for a given $(V,c)$ pair. However, studying the optimal solutions to the LP (\ref{e-lp-o}) would not necessarily give us any insights into the {\em structure} of optimal contests, since we are interested in understanding what contest designs $\M\Mv$ achieve the best possible outcome $p$ for given cost $c$, rather than the contests that have the lowest $c$ for a given $p$.

\subsection{Optimal contests are simple} 
\label{s-simple}
We will now use the LP formulation (\ref{e-lp-o}) to derive the structure of optimal contests. Our main result is that optimal contests are {\em simple}: for every value of total prize $V$ and participation cost $c$, there is an optimal contest which awards $j$ equal prizes of value $V/j$ each to some top $j$ ranks, and a prize of $0$ to all ranks $j+1$ and below. 
% \textcolor{red}{Say something about how striking this is.} 
%
To prove this, we reformulate the LP (\ref{e-lp-o}) by changing variables. For $j = 1, \ldots, n$,  let 
$
w_j = j(v_j - v_{j+1}), 
$ 
where we define $v_{n+1} = 0$. The constraint on the total allocated prize can be written as 
\begin{equation} \label{e-wsum}
V = \sum_{j=1}^n v_j = (v_1 - v_2) + 2 (v_2 - v_3) + 3(v_3 - v_4) + \ldots + n(v_n - v_{n+1}) = w_1 + \ldots + w_n.
\end{equation}
The nonnegativity and monotonicity constraints on the prizes $v_j$,  that is, $v_j \geq v_{j+1}$ for $j = 1, \ldots, n-1$ and $v_n \geq 0$, simply translate to $w_j \geq 0$ for $j = 1, \ldots, n$. 
Finally, we rewrite the objective in terms of the $w_j$: 
\BEAS 
\sum_{j=1}^n v_j \tbinom{n-1}{j-1} p^{j-1} (1-p)^{n-j} &=& \sum_{j=1}^n \left(\sum_{k=j}^n \tfrac{w_k}{k}\right) \tbinom{n-1}{j-1} p^{j-1}(1-p)^{n-j} \\
& = & \sum_{j=1}^n w_j \left( \tfrac{1}{j}\sum_{k=1}^j \tbinom{n-1}{k-1} p^{k-1} (1-p)^{n-k}\right).\\ 
\EEAS
We can therefore rewrite the LP (\ref{e-lp-o}) in the variables $(w_1, \ldots, w_n)$ as follows: 
\BEQ
\begin{array}{cl}
\label{e-lp-o2}
\mbox{maximize}_{\{w_j\}}  &\sum_{j=1}^n w_j \left( \frac{1}{j}\sum_{k=1}^j {n-1 \choose k-1} p^{k-1} (1-p)^{n-k}\right)\\ 
\mbox{s.t.} &\sum_{j=1}^n w_j = V, \\
& w_j \geq 0, \quad j = 1, \ldots n. 
\end{array}
\EEQ

An optimal solution to this LP is set $w_{j^*} = V$ for the index with the largest coefficient multiplying $w_j$ in the objective, \ie, the $w_j$ corresponding to 
\BEQ 
\label{e-optw}
j^* = \argmax_{1 \le j \le n} 
\left\{ \tfrac{1}{j}\sum_{k=1}^j \tbinom{n-1}{k-1} p^{k-1}(1-p)^{n-k} 
\right\},
\EEQ
and $w_j = 0$ for all $j \neq j^*$. We note here that there are only finitely
many values of $p$ for which the $\argmax$ is non-unique, because if
the maximum is achieved at two different values $j,j'$ then this implies
that $p$ satisfies a non-trivial polynomial equation, i.e.
\[
  \tfrac{1}{j} \sum_{k=1}^j \tbinom{n-1}{k-1} p^{k-1}(1-p)^{n-k} =
  \tfrac{1}{j'} \sum_{\ell=1}^{j'} \tbinom{n-1}{\ell-1} p^{\ell-1}(1-p)^{n-\ell}.
\]

What contest structure, \ie, actual awarded prizes $\Mv$ does this LP solution correspond to? Recall our transformation of variables $w_j = j(v_j - v_{j+1})$. With $w_{j^*} = V$ and the remaining $w_j$ set to $0$, the prizes $v_j$ in an optimal contest are then: 
\BEAS
v_j = \tfrac{V}{j^*}, ~1 \leq j\leq  j^*, \quad \mbox{ and } \quad 
v_{j} = 0,~ j^* < j \leq n. 
\EEAS 
Recall that the equation~\eqref{e-optw} defining $j^*$ has a 
unique solution apart from a finite number of exceptional values of $p$.
These exceptions correspond, under the strictly monotone function $c^*(p)$, 
to a finite number of exceptional values of $c$; for all other $c$ 
there is a unique optimal contest and it is a simple contest.
We summarize this result in the following theorem. 

\begin{theorem}
\label{t-ocstruct} 
For every value of total prize $V$ and participation cost $c$, there is an optimal contest which awards equal prizes $V/j^*$ to some top $j^*$ ranks, and a prize of $0$ to all ranks $j^*+1$ and below. Apart from finitely many exceptional 
values of $c$, this is the unique optimal contest.
\end{theorem}

% TDOO:
% - write results about WTA and large-scale asymptotics, without proofs
% - chop to 18 pages
% - fix appendix? later.
% - change mentions of appendix to mentions of full version.

\subsection{Understanding the behavior of the optimal contest} 
\label{sec:compstat}

The result in Theorem \ref{t-ocstruct} illustrates the {\em structure} of optimal contests, namely that there are only two levels of prizes---equal prizes to some top $j^*$ contestants, and a prize of $0$ to all lower-ranked contestants. 
This description of the optimal contest is complete except for one crucial piece of information: what is the value of $j^*$? In this subsection, we seek to understand how the total number of prizes in the optimal contest, $j^*$, varies with the parameters of the contest: the total prize $V$, participation cost $c$, and number of potential contestants, $n$.

A basic question regarding the optimal number of prizes, $j^*$, concerns
comparative statics: is $j^*$ monotonic as a function of participation
cost? It is intuitively clear that as the cost of participation, $c$,
increases, the {\em ex ante} probability of an individual agent 
participating in the equilibrium of the optimal contest, $p^*(c)$, 
must decrease; this intuition was confirmed rigorously in \S \ref{s-lp}.
On the other hand, the monotonicity of
$j^*$ as a function of $c$ is much less apparent:
as the cost of participation increases, should the
contest designer offset the cost increase by increasing
the magnitude of prizes (which necessitates reducing the
total number of prizes offered, and thus the probability
of winning a prize) 
or by increasing the total number of prizes
(which increases the probability of winning a prize but
reduces the magnitude of each prize)?
The following proposition implies, among other things, that an
optimal contest designer responds to increasing participation
costs by offering the same or smaller number of prizes.
All proofs in this section are deferred to Appendix~\ref{s-compstat-proofs}.

\begin{proposition} \label{p-compstat}
Fix a total prize, $V$, and number of potential contestants, $n$.
Let $\M^j$ denote a simple contest that awards equals prizes $V/j$
to the top $j$ ranks. 
%\begin{compactenum}
\BNUM
\item \label{i-compstat1}
There is a decreasing sequence
$V = c_1 > c_2 > \cdots >  c_n > c_{n+1}=0$, such
that for any participation cost $c < V$, the simple contest
$\M^j$ is optimal if and only if $c_{j+1} \leq c \leq c_j$.
\item \label{i-compstat2}
For $1 < j \leq n$, the equation $c_{\M^j}(p) = c_{\M^{j-1}}(p)$
has a unique solution $p_j$ in the interval $(0,1)$, and
$c_j = c_{\M^j}(p_j) = c_{\M^{j-1}}(p_j)$.
\ENUM
%\end{compactenum}
\end{proposition}
%% \begin{proposition} \label{p-compstat}
%% Fix a total prize, $V$, and number of potential contestants, $n$.
%% Let $\M^j$ denote a simple contest that awards equals prizes $V/j$
%% to the top $j$ ranks. 
%% \begin{compactenum}
%% \item \label{i-compstat1}
%% There is an increasing sequence
%% $0 = p_1 < p_2 < \cdots <  p_n < p_{n+1}=1$, such
%% that for any participation cost, $c$, the simple contest
%% $\M^j$ is optimal if and only if $p_j \leq p^*(c) \leq p_{j+1}$.
%% \item \label{i-compstat2}
%% The number of prizes offered by the optimal contest is
%% non-increasing with $c$.
%% \item \label{i-compstat3}
%% For $1 < j \leq n$, $p_j$ is the unique value
%% in $(0,1)$ that satisfies $c_{\M^j}(p_j) = c_{\M^{j-1}}(p_j)$.
%% \end{compactenum}
%% \end{proposition}
Proposition~\ref{p-compstat} allows us to draw a number of
quantitative conclusions about $j^*$, the number of prizes
offered by the optimal contest. The following proposition
identifies a precise 
criterion for optimality of the winner-take-all contest $\M^1$;
for every fixed $j$ 
it identifies an asymptotic formula for $p_j$ as $n \to \infty$,
yielding a criterion for optimality of $\M^j$ that holds for
all sufficiently large population sizes; and finally, it 
presents an asymptotic estimate of the number of prizes 
offered by the optimal contest, and the expected number of
participants in equilibrium, in terms of the ratio $\vc = V/c$.
We call this ratio the {\em scale} of the contest, because
$\vc$ is clearly 
an upper bound on the number of prizes that can be awarded, if each
prize must be at least large enough to cover agents'
participation costs. Proposition~\ref{p-quant} shows
that both the number of prizes awarded and the expected 
number of participants are asymptotic to $\vc$ minus 
a lower-order term. In other words, as the total prize
offered grows large compared each individual agent's participation
cost, the equilibrium of the optimal contest asymptotically
approaches the ideal of ``perfect coordination'',
in which each of the top
$\lfloor V/c \rfloor$ agents (ranked by quality) is asked
to participate and paid just enough to cover their participation cost,
despite the total lack of coordination among agents.

\begin{proposition} \label{p-quant}
Fix $V,c$, and let $j^{*,n}$ denote the number of prizes
offered by the optimal contest with population size $n$.
Let $p^{*,n}$ denote the probability that an individual agent
participates in the equilibrium of that contest, and let
$\ld^{*,n} = p^{*,n} \!\cdot\! n$ denote the expected number
of participants in that equilibrium.
\BNUM
%\begin{compactenum}
\item \label{i-quant-wta}
The winner-take-all contest is optimal if and only if
$\frac{V}{c} \leq \left( 1 + \frac{1}{n-1} \right)^{n-1}$. In particular,
for $\frac{V}{c} > e$ a winner-take-all contest is never optimal, whereas
for $\frac{V}{c} < e$ it is optimal whenever the population size, $n$,
is sufficiently large.
\item  \label{i-quant-limit}
As $n \to \infty$, the sequence
$\ld^{*,n}$ converges to the unique $\ld^* > 0$ satisfying
\BEQ \label{e-ldstar}
\max_{1 \leq j \leq V/c} \left\{ \tfrac{V}{j}  \sum_{k=0}^{j-1}
\tfrac{ e^{-\ld} \ld^{k} }{k!} \right\}  = c.
\EEQ
Excluding a countable set of values for $c$, the maximum
in~\eqref{e-ldstar} is achieved at a unique value $j^*$,
and the sequence $j^{*,n}$ converges to $j^*$ as $n \to \infty$.
\item \label{i-quant-asymp}
Letting $\vc = V/c$, for any $n > 2 \vc$ 
the values of $\ld^{*,n}$ and $j^{*,n}$
satisfy 
\BEQ \label{e-asymptotic}
\ld^{*,n} = \vc - \Theta(\sqrt{\vc \log \vc}), \qquad
j^{*,n} = \vc - \Theta(\sqrt{\vc / \log \vc}).
\EEQ
%\end{compactenum}
\ENUM
\end{proposition}

\section{Arbitrary quality-cost distributions} 
\label{s-hetero}
We now proceed to settings where agents have heterogeneous costs $c_i$ in addition to differing output qualities $q_i$, modeled via a general joint distribution $F(q,c)$ on agents' costs and output qualities. Throughout this section, we assume that $F(q,c)$ is absolutely continuous with respect to Lebesgue measure, \ie, every measure-zero set of $(q,c)$ pairs also has zero probability under the distribution $F(q,c)$.
%\tr{continuous, no point masses, finite support}

When output qualities and costs can have arbitrary correlations, much of the structure we saw in the previous section vanishes. Specifically, equilibria no longer need possess a threshold structure, where all agents whose output quality is greater than a threshold participate and the remaining agents do not. (For example, a very high-ability agent producing very high quality might also have a prohibitively high cost preventing entry, whereas a lower-ability agent might still find it profitable to participate because of an adequately low participation cost $c_i$.) 
The absence of such a threshold equilibrium structure under heterogeneous costs means that unlike with homogeneous costs, there may be no single contest $\M$ that is optimal for all increasing objective functions $f$. {(See % Appendix~\ref{s-counterex} 
\S \ref{s-counterex} for a simple example that illustrates this point; we present a distribution for which there is no contest $\M$ that simultaneously optimizes the sum objective and the max objective.)} Specifying an optimal contest, therefore, involves also choosing the objective that the mechanism designer wants to maximize. 

Motivated by the fact that many contests are aimed at obtaining one best innovation or solution to a problem, or one best design, and so on, a significant fraction of the literature on contest design has focused on the max objective, \ie, the expected quality of the highest-ranked submission. In keeping with this motivation and literature, we focus on the max objective in this section (although we note  that a number of other objectives,  such as the sum or more generally, the sum of the top $k$ outputs, have also been used in prior work, and are potentially well-motivated in the settings we consider as well). Given a total prize $V$, and  $n$ agents with types $(q_i,c_i)$ drawn from $F$, we will therefore define an {\em optimal contest} to mean any mechanism $\M=\M\Mv$ such that some symmetric pure-strategy equilibrium of $\M$ achieves the maximum $E[\max\{q_i~|~i \in \spart(\strat^*(\M))\}]$ over all symmetric pure-strategy equilibria $\strat^*(\M)$ of all contests $\M$ with $\sum v_j = V$, where the set $\spart(\strat)$ is as defined in \S \ref{s-model}.\footnote{As before, we restrict our attention to symmetric equilibria and henceforth omit the modifier ``symmetric.''}

With general type distributions $F$, an agent's type is necessarily two-dimensional and cannot be characterized by a single number. This is in contrast with \S \ref{s-homo}, where agents had homogeneous costs (an agent's type is essentially just her output quality $q_i$), as well as models with strategic effort choices in the prior literature where an agent's type is typically either her ability, or her cost per unit effort. With such arbitrary two-dimensional types, an equilibrium analysis that characterizes equilibrium outcomes in an arbitrary rank-order mechanism appears to be fairly challenging, which means that the standard approach to solving optimal contest design problems---first explicitly solve for equilibrium outcomes for general rank-order mechanisms $\M\Mv$, and then optimize over the choice of feasible $\Mv$ to maximize the objective achieved by a mechanism in equilibrium---cannot be directly applied here. 

%In addition, the following example shows that when the correlations between cost and ability are arbitrary, %\tr{some example to make us respectable.} 

%\tr{EXAMPLE GOES HERE.}

Rather than derive the optimal mechanism, therefore, we ask how well {\em simple} mechanisms
---mechanisms that award equal prizes $V/j$ to the highest-ranked $j$ agents, for some $j$---
can perform, in terms of the expectation of the highest-quality output in equilibrium relative to an optimal mechanism. In this section, we will show that the winner-take-all mechanism (corresponding to $j=1$), which awards the entire prize budget $V$ to the participant with the highest quality output, is a $3$-approximation to the optimal mechanism. 
%has expected output (in any equilibrium) which is at most a factor $3$ smaller than the expected output from the (best equilibrium of) an optimal mechanism. 
% (this, in addition to Example \ref{e-xxx}, provides another reason why deriving the exactly optimal contest might be difficult for general distributions $\rho$ without imposing any further structure). 

\paragraph{Existence and uniqueness of pure strategy equilibria}
Before delving into the analysis of symmetric equilibria
of contests under arbitrary joint distributions $F$,
we pause here to mention some basic facts which ensure that
the analysis is meaningful. Proofs of these facts are given
in Appendix~\ref{s-exist-proofs}.
%% In this section we present the following basic results about
%% equilibria of contests in our model. 
First, for every contest
$\M$ there exists a symmetric mixed-strategy equilibrium. Second, 
all mixed-strategy equilibria of $\M$ are ``pure almost
everywhere,'' in the sense that the set of types that 
randomize in equilibrium has measure zero.
Third, the symmetric equilibrium of $\M$ is unique
up to an indeterminacy on a measure-zero set of types $(q,c)$.

\subsection{Sub-equilibrium lemma}
\label{s-subeq}

As discussed above, solving for the equilibrium of a mechanism under arbitrary joint distributions $F$ is challenging. To circumvent the need to solve for equilibria, we instead relate the expected outcome in an equilibrium of the winner-take-all mechanism $\WTA$ to that in an optimal mechanism by identifying and analyzing {\em sub-equilibrium} strategy profiles in $\WTA$, as defined below. 

\begin{definition} \label{def:semieq}
A strategy profile $(\strat_1,\ldots,\strat_n)$ 
is a {\em sub-equilibrium} of a given contest
if it satisfies {\em ex interim} individual rationality: for any player $i$
and type $(q_i,c_i)$ such that $\strat_i(q_i,c_i)=1$, the expected payoff of 
player $i$ from participating, conditional on having type $(q_i,c_i)$,
is greater than or equal to $c_i$ when $n-1$ remaining players with types drawn from $F(q,c)$ play according to $\strat$. (Note that individual rationality vacuously holds for non-participants, \ie, agents with $\strat_i(q_i,c_i) = 0$, since non-participation yields zero utility and is therefore always individually rational.) We say that a pure strategy $\strat$ is a sub-equilibrium if the corresponding symmetric strategy profile $\vec{\strat}$ with $\strat_i = \strat$ for all $i$ is a sub-equilibrium.
\end{definition}

\iffalse 
\begin{definition} \label{def:max-notation}
Suppose $P$ is any $\{0,1\}$-valued function on $\reals_+^2$.
% and $\rho(a,c)$ is a joint distribution on $\reals_+^2$. 
The function $\rvm(P) : \reals_+^{2n} \to \reals_+$
is defined to be the function that maps a tuple
$(a_1,c_1,a_2,c_2,\ldots,a_n,c_n)$ to the number
  $\max \{ a_i P(a_i,c_i) \mid i=1,\ldots,n \}$.
If one makes $\reals_+^2$ into a probability space under the 
joint distribution $\rho(a,c)$, then $\reals_+^{2n}$ becomes 
a probability space under the product measure $\rho^n$ and
we will be interpreting $\rvm(P)$ as a random variable on this
probability space.

We will abuse notation and refer to $\rvm(P)$ also when $P$ is 
a subset of $\reals_+^2$ or a Boolean predicate of variables $a,c$. 
In the former
case, this amounts to ignoring the distinction between a set
and its characteristic function; in the latter case, it amounts
to ignoring the distinction between a Boolean predicate and the
function that maps $(a,c)$ to $1$ if the predicate is true, $0$
if it is false.
\end{definition}
\fi 
A sub-equilibrium is a strategy profile that is equilibrium-like, but not quite an equilibrium: all agents who do participate derive non-negative utility, so that their action choice is indeed a best response, but there might be non-participating agents who would derive non-negative expected utility from participation given other agents' strategies, so that these agents' choices do not constitute a best response. A sub-equilibrium is therefore a strategy profile with `too little' participation, where all {\em participants} are playing a best response, but non-participants might have a profitable deviation. 

The interpretation of sub-equilibria as strategy profiles with `too little' participation suggests that if $\sstrat$ is a sub-equilibrium and
$\estrat$ is an equilibrium, then the quality of an agent's output (treating
non-participation as zero quality) in $\estrat$ should stochastically dominate 
the agent's output quality in $\sstrat$. Lemma~\ref{l-se} below confirms this
intuition. Note, however, that it is not obvious that this intuition 
should be valid: some agent types who choose non-participation
in $\sstrat$ may choose to participate in $\estrat$, but 
the participation of these types may adequately decrease the probability of winning for other agent types to cause them to flip their choice from participation to
non-participation, as a rational response to the increased 
competition. That these effects should lead, on net, to
stochastic dominance appears {\em a priori} to be far from evident.

%% Recall from \S \ref{s-model} the definition $\spart(\strat)$ of the (random) set of participants when all agents make participation decisions according to strategy profile $\strat$. The following result relates the expected value of the max objective in any equilibrium and sub-equilibrium strategy profiles of the winner-take all mechanism. 

\begin{lemma}[Sub-equilibrium Lemma] \label{l-se}
Fix a joint distribution $F(q,c)$ of types in
an $n$-player contest $\M=\M(v_1,\ldots,v_n)$. 
Assume that $v_1 \neq v_n$ and 
that the marginal distribution of $q$ has no point masses.
If pure strategy $\sstrat$ is any sub-equilibrium and pure strategy $\estrat$
is any equilibrium of $\M$, then the random variable $q \cdot \estrat(q,c)$
stochastically dominates the random variable $q \cdot \sstrat(q,c)$.
% $\max\{a_i|i \in \spart(\estrat)\}$ stochastically dominates the random variable $\max\{a_i|i \in \spart(\sstrat)\}$. 
\end{lemma}

\begin{proof}
Let $F_S, F_E$ denote the cumulative distribution
functions of the random variables $q \cdot \sstrat(q,c)$ and
$q \cdot \estrat(q,c)$, respectively. The lemma asserts that
$F_S(x) \geq F_E(x)$ for all $x$. Our assumption that the
marginal distribution of $q$ has no point masses implies that the
marginal distributions of $q \cdot \sstrat(q,c)$ and $q \cdot \estrat(q,c)$ 
likewise have no point masses and hence $F_S, \, F_E$ are 
continuous functions on the set of non-negative real numbers, 
$\reals_+$. 

If an agent $i$ with type
$(q,c)$ chooses to enter the contest $\M$, and all other 
agents play strategy profile $\sstrat_{-i}$, then the 
probability that a given one of those opponents participates
and ranks above agent $i$ is $1-F_S(q)$. Therefore, if agent $i$
chooses to participate, the
probability that she ranks $j^{\mathrm{th}}$ is equal to 
$\binom{n-1}{j-1} (1-F_S(q))^{j-1} F_S(q)^{n-j}$ and her
expected prize is equal to $\sum_{j=1}^{n} v_j \binom{n-1}{j-1}
(1-F_S(q))^{j-1} F_S(q)^{n-j}$. Recalling the definition of
the function $c_{\M}$ from \S \ref{s-homo}, we see that an
agent with type $(q,c)$ who chooses to participate when the
opponents play strategy profile $\sstrat_{-i}$ will gain
an expected reward of $c_{\M}(1-F_S(q))$. 

To prove that $F_S(x) \geq F_E(x)$ for all $x$, we argue by
contradiction. Assume that $F_S(x_0) < F_E(x_0)$. Since
the function $G = F_E - F_S$ is continuous on $\reals_+$, 
there is an open interval $I$
containing $x_0$ such that $G$ is strictly positive
on $I \cap \reals_+$.
The union of all such open intervals is itself a non-empty
open interval $(\underline{x},\overline{x})$; here we allow 
$\underline{x} = -\infty$ or $\overline{x} = +\infty$, or both.
We know that $\overline{x}>0$ because
$\overline{x} > x_0$.
In $0 < \overline{x} < \infty$, then the continuity of
$G$ implies that $G(\overline{x})=0$ and that
\begin{equation} \label{eq:lem:semieq.0}
\lim_{x \to \overline{x}} G(x) = 0.
\end{equation}
If $\overline{x} = \infty$
then $F_E(x)$ and $F_S(x)$ both converge to 1 as 
$x \to \overline{x}$, and so again~\eqref{eq:lem:semieq.0} holds.

For any $(q,c) \in \reals_+^2$ 
such that $\sstrat(q,c)=1$ and $\underline{x} < q <
\overline{x}$, we have
\begin{equation} \label{eq:lem:semieq.1}
    c \leq c_{\M}(1-F_S(q)) < c_{\M}(1-F_E(q)),
\end{equation}
where the first inequality expresses the individual
rationality constraint in the definition of sub-equilibrium,
and the second inequality holds because $F_E- F_S$
is positive on interval $(\underline{x},\overline{x})$
and the function $c_{\M}(1-y)$ is strictly increasing in $y$
(by Lemma~\ref{l-qc1to1}). Since $\estrat$ is an equilibrium, 
inequality~\eqref{eq:lem:semieq.1} implies
that $\estrat(q,c)=1$, since $c_{\M}(1-F_E(q))$ is the expected prize to an agent with quality $q$ in the strategy profile $\estrat$. Thus, we have shown that 
$\sstrat(q,c)=1$ implies $\estrat(q,c)=1$ when
$\underline{x} < q < \overline{x}$. Now, if $\underline{x} < x_0 < x_1 < \overline{x}$, we have
\begin{align*} 
  F_S(x_1)-F_S(x_0) &= 
  \Pr_{(q,c)} ( x_0 < q < x_1, \; \sstrat(q,c)=1 ) \\ 
    &\leq
  \Pr_{(q,c)} ( x_0 < q < x_1, \; \estrat(q,c)=1 ) 
    =
  F_E(x_1)-F_E(x_0)
\end{align*}
%% \begin{equation} \label{eq:lem:semieq.3}
%% F_\equil(x_0) - F_\strat(x_0) \leq F_\equil(x_1) - F_\equil(x_1).
%% \end{equation}
and by rearranging terms on the left and right sides we find that
$G(x) \geq G(x_0) > 0$ for all $x \in (x_0,\overline{x})$.
This means that $\lim_{x \to \overline{x}} G(x)$ cannot be $0$, contradicting~\eqref{eq:lem:semieq.0} and completing the proof.
\end{proof}

Recall from \S \ref{s-model} that $\spart(\strat)$ is the (random) set of participants when all agents make participation decisions according to strategy profile $\strat$.
The following immediate corollary of Lemma~\ref{l-se} will be needed in
the sequel. 
\begin{corollary} \label{c-se-max}
Fix a joint distribution $F(q,c)$ of types in
an $n$-player winner-take-all contest $\WTA$.
Assume that the marginal distribution of $q$ has no point masses.
If pure strategy $\sstrat$ is any sub-equilibrium and pure strategy $\estrat$
is any equilibrium of $\M$, then 
$\max\{q_i|i \in \spart(\estrat)\}$ stochastically dominates the random variable $\max\{q_i|i \in \spart(\sstrat)\}$. 
\end{corollary}

\subsection{Winner-take-all is a $3$-approximation of the optimal contest}
\label{s-wta-approx}

We will now sketch the proof that the expected value of the maximum quality entry in any pure-strategy equilibrium of the winner-take-all contest $\WTA$ is no smaller than one-third of the expected maximum quality in the best equilibrium of an optimal contest $\M^*$. 
% Once again, the detailed proof is given in the full version of the paper, while proofs of some steps are omitted here for space reasons.

Consider an optimal contest $\M^*$ and let $\estrat$ denote its symmetric equilibrium (recall % from Lemma~\ref{l-uniq} 
that $\estrat$ is unique). We use $\eopt = \spart(\estrat(\M^*))$ to denote the random set of participant types in the equilibrium strategy profile $\estrat(\M^*)$ of $\M^*$, 
%$
%\eopt = \spart(\estrat(\M^*)).
%$ 
and let the random variable $m(\eopt) =  \max\{q_i~|~i \in \eopt\}$ denote the maximum quality in a particular instantiation of the random set $\eopt$ 
%$
%m(\eopt) = \max\{q_i~|~i \in \eopt\}. 
%$ 
(recall that the principal's objective is to maximize the expected highest quality). 
Note that 
\[ 
m(\eopt) \leq \max \left\{ q_i~|~i \in \eopt; c_i \leq \tfrac{V}{2} \right\} + \max \left\{q_i~|~i \in \eopt; c_i > \tfrac{V}{2} \right\}, 
\] 
since the highest-ability agent that participates in $\M^*$ for any given random instantiation of $\eopt$ has cost either less than, equal to, or greater than $V/2$. (We define the max of an empty set to be zero.)

We will use $\opt$ to denote the expected value of $m(\eopt)$ over the random draws of the $n$ types $(q,c)$. % from $F$. %: $ \opt = E[m(\eopt)]$. 
Since the inequality above holds pointwise, it also holds in expectation: 
\BEQ 
\label{e-3-ineq1}
\opt =  E[m(\eopt)] \leq E \left[ \max \left\{
  q_i~|~i \in \eopt; c_i \leq \tfrac{V}{2} \right\} \right] 
        + E \left[ \max \left\{
  q_i~|~i \in \eopt; c_i > \tfrac{V}{2} \right\} \right]. 
\EEQ
Now consider the winner-take-all mechanism $\WTA$. Denote the expected value of the maximum output $q_i$ amongst the set of agents participating in the equilibrium $\estrat(\WTA)$ of the winner-take-all mechanism by $\wta$, \ie,
$$
\wta = E[\max\{q_i|i \in \spart(\estrat(\WTA))\}]. 
$$
We will show that $3 \!\cdot\! \wta \geq \opt$ by showing that the first term is (\ref{e-3-ineq1}) is no larger than twice $\wta$ and the second term in (\ref{e-3-ineq1}) is no larger than $\wta$. That is, imagine the expected quality of the best submission in the optimal contest as `split' across two subpopulations---the low-cost subpopulation (with $c_i \leq V/2$), and the high-cost subpopulation (with $c_i > V/2$). We show that the winner-take-all mechanism does at least as well as the optimal mechanism `restricted' to the high-cost subpopulation, and no worse than half as badly as the optimal mechanism `restricted' to the low-cost subpopulation. Note that the nature of these comparisons should not be entirely unexpected from our results in \S \ref{s-homo}---in a homogeneous-cost model, the winner-take-all mechanism is suboptimal when participation costs are small relative to the total prize, but optimal when the cost $c$ is large enough compared to $V$, suggesting that $\WTA$ is likely to `lose' relative to the performance of an optimal contest with a low-cost subpopulation, but not a high-cost subpopulation. The following two inequalities summarize this discussion.
%\begin{lemma} 
\BEA
\label{l-comp1}
\wta & \geq & \frac{1}{2}E[\max\{q_i~|~i \in \eopt; c_i \leq \tfrac{V}{2} \}] \\
\label{l-comp2}
\wta & \geq & E[\max\{q_i~|~i \in \eopt; c_i > \tfrac{V}{2} \}]
\EEA
%\end{lemma}
%% Next we prove the second lemma, 
%% % namely that 
%% relating the expected maximum output 
%% % from agents who choose to participate in 
%% in the equilibrium of $\WTA$ 
%% % is no smaller than 
%% to the expected maximum output from agents who participate in the equilibrium of the optimal mechanism and have  large costs $c_i \geq \frac{V}{2}$. 
%% % , in any given equilibrium of the optimal mechanism $\M^*$:
%\begin{lemma} 
% \label{l-comp2}
% \wta \geq E[\max\{q_i~|~i \in \eopt; c_i > \frac{V}{2} \}]
%\end{lemma}
The proofs of both inequalities are presented in Appendix~\ref{s-hetero-proofs}.
Both proofs rely on Lemma~\ref{l-se}, 
the sub-equilibrium lemma: in both cases, the proof proceeds
by constructing a sub-equilibrium 
of $\WTA$ and showing that the expectation of the maximum quality in
that sub-equilibrium---for brevity, 
the \emph{value} of the sub-equilibrium---is an upper bound 
on the corresponding term 
in~\eqref{e-3-ineq1}, or half of that term in the case 
of Inequality~\eqref{l-comp1}. To prove Inequality~\eqref{l-comp1}, we
use a sub-equilibrium in which agents participate if their
cost is at most $\frac{V}{2}$ and their
quality is at least $\mu$, where $\mu$ is 
chosen so that the probability that
$\max \{q_i \mid c_i \leq \frac{V}{2} \}$
exceeds $\mu$ is exactly $\frac12$. This is
easily seen to be a sub-equilibrium: agents
only participate if their probability of winning
is at least $\frac12$ in which case their expected
prize is at least $\frac{V}{2}$, while their cost
is at most $\frac{V}{2}$, which guarantees a non-negative utility. The proof that the value
of the sub-equilibrium is at least the RHS of % the inequality in 
Inequality~\eqref{l-comp1} is similar to
Samuel-Cahn's~\citeyearpar{S-C} proof of the 
Prophet Inequality. To prove Inequality~\eqref{l-comp2} we
use a different sub-equilibrium, where the set of 
participating agent types equals the set of all
types that participate in the equilibrium of the 
optimal contest {\em and} have cost greater than $\frac{V}{2}$.
The value of this strategy profile is precisely the second
term in~\eqref{e-3-ineq1}; the difficulty here is to prove that the 
strategy profile is indeed a sub-equilibrium.% strategy profile in $\WTA$. 
We do this by showing that any monotone rank-order mechanism
is payoff equivalent to a randomization over simple contests,
\ie, contests which award $j$ equal prizes.
% (\ie, contests which award $j$ equal prizes of size $\frac{V}{j}$); 
This means that for a type $(q,c)$ that participates
in the equilibrium of the optimal contest,
it must be individually rational to participate in at least one simple contest. % On the other hand, an 
An agent with cost greater than $\frac{V}{2}$ cannot find it individually rational to participate in 
any simple contest that awards more than one prize, so for the participating types in our putative sub-equilibrium, participation in a simple contest with one prize---that is,
a winner-take-all contest---must be individually rational.
This shows that the putative sub-equilibrium is actually
a sub-equilibrium and completes the proof. 

We note that neither of these proofs rely on any structural property specific to the optimal contest:  all steps in both proofs remain valid if we instead compare $\WTA$ to an
arbitrary monotone rank-order mechanism respecting the sum constraint ($\sum_{j=1}^n v_j = V$), whether or not it is optimal. In fact, this aspect of our proofs is in some sense inevitable since we have almost no insight into the structure of the optimal contest in this setting with two-dimensional agent types resulting from arbitrary quality-cost distributions; the analysis therefore necessarily relies on properties that hold for all feasible contests. 

%% Note here that neither the arguments in Lemma \ref{l-comp1} nor Lemma \ref{l-comp2} rely on any knowledge about the structure of an {\em optimal} mechanism: the comparisons in both lemmas would apply to {\em any} feasible mechanism, and in particular, to an optimal mechanism. Specifically, Lemma \ref{l-comp1} uses the fact that it is enough to establish a comparison between the expected outcome in $\WTA$ and the expected maximum quality amongst the subset of $n$ agents with draws $(q_i,c_i)$ such that $c_i \leq V/2$;  this argument has nothing to do with the optimal mechanism and only relies on the payoff structure of the winner-take-all mechanism. The second lemma, relating the outcome in $\WTA$ to high-cost agents, appears to use a decomposition of the optimal mechanism into a lottery over simple mechanisms. Note, however, that the validity of this decomposition, and the arguments that follow, hinge only on the feasibility of the mechanism (\ie, that it allocates rewards $(v_1, \ldots, v_n)$ with $\sum_{j=1}^n v_j = V$. The argument therefore applies to {\em any} mechanism, so that the lemma really proves that $\WTA$ produces an outcome that is at least as good as that produced by {\em any} mechanism---and therefore an optimal mechanism---restricted to the high-cost population.  

Inequalities~\eqref{l-comp1} and~\eqref{l-comp2} combine to yield our main
approximation result.
\begin{theorem} \label{t-3approx}
The expected maximum output in the winner-take-all mechanism is at least one-third that in the optimal mechanism: $\opt \leq 3 \!\cdot\! \wta$. 
\end{theorem}
%% OMIT THIS PROOF FROM THE PROCEEDINGS VERSION
\iffalse
\proof From \eqref{e-3-ineq1}, and Lemmas \ref{l-comp1} and \ref{l-comp2}, 
\[
% \BEAS
\opt \leq E[\max\{q_i~|~i \in \eopt; c_i \leq \frac{V}{2} \}] + E[\max\{q_i~|~i \in \eopt; c_i > \frac{V}{2}\}]
\leq  2\wta + \wta 
= 3\wta, 
%\EEAS
\]
where the second inequality uses Lemma \ref{l-comp1} to bound the first term, and Lemma \ref{l-comp2} to bound the second. 
\endproof
\fi

\subsection{Dependence on the objective function}
\label{s-counterex}

We conclude this section with a simple example illustrating
an important difference between the general two-parameter 
setting studied in this section and the homogeneous-cost 
setting in \S \ref{s-homo}. Unlike the case of
homogeneous costs, for a general joint distribution of 
output quality and cost, there may be no single contest
that is optimal for all choices of the increasing objective
function $f$: Example~\ref{ex-obj} below 
presents a joint distribution
$F(q,c)$ such that the optimal contest for the sum objective 
differs from the optimal contest for the max objective. 
%% Second, Example~\ref{ex-simple}
%% illustrates that even when dealing with the max objective, 
%% the structure of the optimal contest is not necessarily
%% simple, \ie it may be optimal to divide the total prize 
%% $V$ into unequal share.

\begin{example} \label{ex-obj}
Suppose that the joint distribution $F(q,c)$ 
is supported on the squares $[1,1+\epsilon] \times [1-\epsilon,1]$
and $[20,21] \times [\frac{9V}{10}-1,\frac{9V}{10}]$, 
with probability density $1/(2 \epsilon^2)$ on the
first square and $1/2$ on the second one. (Thus,
to sample from $F(q,c)$ one first selects among
the two squares with equal probability, then one 
chooses a uniformly random point in the selected
square.) Denote the types $(q,c)$ with $q > 1+\epsilon$
as {\em high types} and those with $q \leq 1+\epsilon$
as {\em low types}.
Assume that $0 < \epsilon \ll 1$ and
$n \gg V \gg 1$.
% {\bf and other assumptions.}

Choose $x \in [20,21]$ to satisfy 
$\Pr_{(q,c) \sim F} (q < x) = (9/10)^{1/n}$.
Equivalently,
$$ \Pr(\max \{q_1,\ldots,q_n\} < x) = \tfrac{9}{10} $$
when $n$ independent types $(q_i,c_i)$ are
sampled from $F$. An agent
whose output quality is greater than $x$ has probability
at least $9/10$ of having the highest ability 
among all agents. Such an agent
gains an expected prize of at least $9V/10$ in a
winner-take-all contest $\WTA$, no matter what strategy
profile the other agents use, and therefore the
equilibrium of $\WTA$ has full participation of
agents with quality greater than $x$. The probability
that at least one of the $n$ agents samples a type
$(q_i,c_i)$ with $q_i>x$ is $\frac{1}{10}$, so the
expectation of the maximum output quality in the
equilibrium of the winner-take-all contest is
at least $\frac{x}{10} > 2$.

In comparison, any contest with $v_1 < \frac{9V}{10}-1$ cannot 
elicit the participation of any high-type agents, 
and hence the expectation
of the max objective, in the equilibrium of such a contest,
is less than $1+\epsilon$. This shows that 
the optimal contest for the max objective must satisfy
$v_1 \geq \frac{9V}{10} - 1$.

Now consider the sum objective. If $\M$ denotes a contest that 
offers $V/2$ equal prizes of size $2$, we claim that in the
equilibrium of $\M$, the expected sum of output qualities is
at least $V/4$. Indeed, high types do not participate in 
equilibrium (the prize is less than their cost) so only the
low types participate. Let $y \in [1,1+\epsilon]$ be 
such that, in expectation, exactly $V/4$ of the low types among
$\{(q_1,c_1),\ldots,(q_n,c_n)\}$ satisfy $q_i > y$. 
By Markov's Inequality, the probability that more than
$V/2$ low types with $q_i > y$ participate is at most
$1/2$, and hence an agent $i$ with $q_i > y$ 
has probability at least $1/2$ of winning a prize
of size $2$. If the agent is of low type, then the
expected prize exceeds the agent's cost, and 
consequently all of  the low-type agents with
$q_i>y$ will participate in equilibrium. The
expected number of such agents is $V/4$ and each
of them contributes an output of quality at least 1,
so that expectation of the sum objective in equilibrium
is at least $V/4$ as claimed.

In comparison, in any contest with $v_1 \geq \frac{9V}{10}-1$,
the total prize awarded to contestants in ranks 2 and higher
is at most $\frac{V}{10}+1$. Define a random variable $Z$ to be
equal to the total cost expended by the agents who participate
and place in ranks 2 and higher. Individual rationality 
dictates that $E[Z] \leq \frac{V}{10} + 1$. However, 
the ratio $q/c$ is at most $\frac{1+\epsilon}{1-\epsilon}$
for all points $(q,c)$ in the support of the type distribution.
Consequently, the expected total quality of the submissions
from agents who place in ranks 2 and higher is bounded above
by $\frac{1+\epsilon}{1-\epsilon} \left( \frac{V}{10} + 1 \right)$.
The top-ranked agent has output quality at most 21, and hence
the expectation of the sum objective in equilibrium is at most
$21 + \frac{1+\epsilon}{1-\epsilon} \left( \frac{V}{10} + 1 \right)$,
which is less than $\frac{V}{4}$ for sufficiently large $V$.

To summarize, we have shown that the optimal contest for the
max objective satisfies $v_1 \geq \frac{9V}{10} -1$
whereas the optimal contest for the sum objective
satisfies $v_1 < \frac{9V}{10} - 1$, so no single
contest is optimal is optimal for both objectives.
\end{example}

\section{Discussion}
%Summary: Theory and practical implications 
In this paper, we explored {\em behavioral} design---the idea that `non-standard' agent behavior might have  consequences for the optimal design of economic environments---via the specific problem of contest design for {\em simple} agents, who strategize only about whether or not to participate in a system rather than also strategizing about the input they provide to it. 

Our results have potential practical implications for the design of contests as well as contest-like environments (such as online communities that hand out virtual rewards to top contributors) where the {\em quality} of a participant's contribution is not the result of strategic deliberation. The structure of optimal contests in homogeneous cost environments suggests, for example, that a website might create stronger incentives for optimal contribution if it awards a (suitably-chosen) number of identical badges to top contributors instead of rewarding top contributors via recognition on a leaderboard, which might yield unequal rewards at  different leaderboard ranks. In settings with heterogeneous costs, as in online crowdsourcing contests where participants may differ both in expertise and in cost (of effort or time) to contribution, our results show that a detail-free winner-take-all mechanism, while not necessarily optimal, is always approximately optimal---a guarantee which does not depend, for example, on the specific form of the cost-quality correlation, or the size of the contestant population, or the available total reward. 

The differing structure of equilibria in the homogeneous and heterogeneous cost settings, as well as the example in \S \ref{s-counterex} showing that the same contest cannot simultaneously (exactly) optimize the max and sum objectives with heterogeneous costs, suggest another interesting practical implication. The designer of an online community in which contribution is perceived as essentially equally costly by all agents does not have to choose whether her objective is to maximize participation (`build a community'), or maximize the quality of the best contribution or the best few contributions (`create excellent content'), or any combination thereof---as we saw in \S \ref{s-homo}, all (increasing) objectives of the designer are maximized by the same reward structure in this setting. In contrast, the designer of a platform where users may have differing costs to participation---such as busy experts with a high time-cost to contribution versus users with lower expertise and lower costs---might have to make a choice about which objective she would like her platform to optimize, and  design her choice of rewards accordingly.

\subsection{Further directions}
%Our work thus far provides an indication that agent behavior can indeed have consequences for design as well as lead to interesting new theoretical questions, but also raises more questions than it answers, at various levels.

Our results answer the question of how to design optimal contests for simple agents---exactly, under  homogeneous costs, and approximately, for hetereogeneous costs, and suggest that `non-standard' agent behavior can indeed have consequences for optimal design, as well as lead to interesting new theoretical questions. At the same time, our work also raises more questions than it answers---at various levels---and points to several avenues for further work. \\

\noindent {\bf Robustness. } An intriguing family of open questions, in the specific setting of optimal contest design for simple agents, centers around the theme of robustness or simultaneous approximate optimality. 
\BIT 
\item Robustness to {\em objective}: We saw that the winner-take-all mechanism, which awards the entire prize of $V$ to the top-ranked agent, is a 3-approximation for the max objective in the setting with arbitrary cost-quality correlations. Are there simple, detail-free, contests that are also simultaneously approximately optimal for other objectives that the designer might have, such as the sum of qualities? 

\item Robustness to {\em parameters}: Consider the setting with homogeneous costs, where we showed that optimal contests are {\em simple}, \ie, award equal prizes to the $k$ top-ranked agents for some $k$. A principal without complete knowledge of the population parameters might choose the wrong number of prizes. How suboptimal can the `wrong simple contest' be, and is there a contest which is nearly optimal for all population parameters\footnote{Note that the answer to this last question might depend on the objective that the designer wants to optimize.}?

%% An important issue to address, both here and in the context of the preceding question, is what the right notion of approximation itself is. Multiplicative bounds on the expected value of the objective function, as in our Theorem~\ref{t-3approx}, are meaningful in situations where the quality of a participant's contribution, $q_i$, can meaningfully be interpreted as a {\em cardinal} value. In settings where only ordinal quality comparisons are meaningful, it is more appropriate to use other notions of approximation. As a first step toward obtaining approximation results in ordinal settings, one could seek  approximation guarantees for mechanisms that hold simultaneously with respect to all monotonic reparameterizations of the quality scale. (Our result on the WTA contest already has this property.)

%% , for example those that pertain to the expected number of contributions {exceeding a specified quality threshold (AG: Doesn't this require a cardinal value?)}.
%%  An immediate question is whether the sub-equilibrium lemma can be used to facilitate such a result, which might provide a pathway to designing almost-optimal mechanisms for other objectives. 

\item Robustness to {\em population}: The model in our paper assumes that all agents are simple, whereas the past literature on contest design involved models where all agents make strategic effort choices. What if the population consists of a {\em heterogeneous} mix of agent types---those who make participation-only choices, and those who make strategic effort choices as well? Are there  mechanisms that are robust to such heterogeneity in the {\em sets of alternatives} from which agents choose actions? 
\EIT 
Exploring the robustness and suboptimality of simple contests is a practically relevant, although theoretically challenging, direction for further work, since a principal designing a real-world contest is more likely than not to suffer from uncertainty regarding population parameters, as well as have agents of differing types in her population. \\

\noindent{\bf Tournament design. } A second question concerns optimal {\em tournament} design for simple agents. In this paper, we restricted ourselves to choosing amongst rank-order mechanisms where an agent's reward depends only on her rank amongst all competitors. In general, of course, competitions need not be restricted to such reward structures: indeed, many real-life contests do split agents into subdivisions or categories, and award category-specific as well as overall prizes. Analogous to the reward allocation problems analyzed in \cite{MS06} (which build on the techniques developed for optimal contest design in \cite{MS01}) for agents who make strategic effort choices, what are optimal {\em tournament} designs for simple agents? 

%General 
More generally, the notion of simple agents---agents who reason strategically about {\em whether}, but not about {\em how}, to participate in a system---seems to be a natural one in many scenarios. In addition to our running example of online content contribution platforms, settings such as crowdfunding websites (where investors might decide whether or not to contribute a recommended sum, rather than strategize over how much to contribute), online ratings, or voting (on content, or in participatory democracies) might elicit such simple behavior from at least some of the user base. How does the design of optimal mechanisms change when agents make participation-only choices? 

% More generally, a model with simple agents who do reason about whether or not it is worth their while to participate in a system, but who do not strategize carefully about what input they provide to the system if participating, seems natural in a variety of scenarios---the benefits from participation might be intrinsic, or be too small to justify the effort of carefully strategizing over choices, or simply because participation is the only costly choice and users have little reason to strategize over their input. In addition to our running example of online content contribution platforms, settings such as `small' auctions for items with very low values (for examples, individual instances of ad auctions with very small bids), contribution on crowdfunding websites where investors might decide whether or not to contribute some fixed amount but not strategize over how much, online matching markets such as dating or item swaps, and online rating or voting---on content, for example, or in participatory democracies---might elicit such simple behavior from at least some of their user base. How does the design of optimal mechanisms change when agents make participation-only choices? 
%---what are the implications for design, as well as for new theoretical techniques? 

While one might expect that a model with simpler agents ought to result in simpler mechanism design questions, we have seen in \S \ref{s-hetero} that restricting agents to make discrete (binary) choices rather than selecting a level of effort from a continuous interval leads to qualitatively different equilibria and requires qualitatively different analysis techniques. This suggests that designing for simple agents---a problem that might arise in a variety of economic settings---may require the development of new theoretical tools and techniques.

Finally, a population of simple agents is one particular instance of a model for behavioral mechanism design, aimed at more effectively incentivizing desired outcomes in an environment by using more accurate models of agent behavior in that environment. A range of other deviations from idealized, or standard, models of behavior have also been discussed in the literature, ranging from workers' preferences in online crowdsourcing labor markets~\cite{HC10}, to a vast variety of  biases and fallacies documented in the growing behavioral economics literature. The problems of developing better models of agent behavior in economic environments---an empirical and experimental question---and that of optimally designing mechanisms or markets for the real populations they are targeted at---a theoretical question---are closely intertwined, and constitute a promising and exciting direction for future research. 

\iffalse
While each of these corresponds to a different specific model of behavior, a broader general question relates to {\em heterogeneity} in the population: what if the population of agents is heterogeneous---in terms of how it behaves? For example, what if the population consists a mix of simple agents making participation-only choices, and `standard' agents who make choices over a `complete' choice set (this might occur, for instance, in ad auctions where there are a number of small mom-and-pop advertisers who might not have the wherewithal to strategize carefully about their bids, and a few large advertisers with large campaigns who do reason strategically about their bids)? We have seen already that optimal mechanisms for populations consisting only of agents of one kind can differ depending on which kind. Do there even exist mechanisms  and if yes, what are such mechanisms? The problem of the existence and design of {\em robust} mechanisms that are simultaneously approximately optimal for various mixes of agents in the population is an exciting direction for further work. 
\fi

\subsection*{Acknowledgments}
Robert Kleinberg acknowledges support from NSF award AF-0910940, 
a Microsoft Research New Faculty Fellowship, and a Google Research Grant.

\bibliography{contests}
%\bibliography{participation_BK}

 \appendix

\section{Proofs from Section~\ref{s-homo}}
\label{s-homo-proofs}

This section provides proofs deferred
from \S \ref{s-homo}.
We begin by restating and proving Proposition~\ref{p-eq-homo}.
\newtheorem*{p-eq-homo}{Proposition \ref{p-eq-homo}}
\begin{p-eq-homo} 
For any monotone rank-order mechanism $\M$, there is a symmetric equilibrium in threshold strategies where every agent with quality $q_i \geq \at$ participates, and every agent with quality $q_i < \at$ does not participate; the equilibrium threshold $\at$ satisfies $c_{\M}(1-F(\at)) = c$.
This is the unique symmetric equilibrium of $\M$.
\end{p-eq-homo}
\proof
Let $\strat$ be any symmetric pure-strategy equilibrium of $\M$ and let
$q_0 = \inf \{q \mid \strat(q,c)=1\}$. By the definition of
$q_0$, for any $q > q_0$ there is some $q_1 \leq q$ such that 
$\strat(q_1,c)=1$. Since $\strat$ is an equilibrium, this implies 
that an agent with quality $q_1$ earns an expected prize 
greater than or equal to $c$ in a contest with $n-1$ potential opponents each playing strategy $\strat$. Since $\M$ is 
monotone, the expected prize given to an agent is a
non-decreasing function of her output quality, 
so that an agent
with quality $q > q_1$ earns an expected prize greater
than or equal to her cost: since ties are broken in favor of participation, this means
that $\strat(q,c)=1$; and this holds
for all $q > q_0$. By the definition of $q_0$, $\strat(q,c)=0$ for all $q < q_0$; 
finally, our rule that ties are broken in favor of
participation, combined with our assumption that 
the distribution of $q$ has no point-mass at $q_0$,
ensures that $\strat(q_0,c)=1$. Hence, $\strat$ is 
a threshold-strategy equilibrium with threshold
$\at = q_0$.

Next we characterize the equilibrium threshold $\at$, and argue that it is the solution to $c_{\M}(1-F(\at))=c$. Note that when all agents play strategy 
$\strat$, $1-F(\at)$ is the probability that a
given agent participates and contributes an output
with quality greater than or equal to $\at$. 
Recalling our definition of 
$c_{\M}$, and noting that when agents play according
to $\strat$ they \emph{never} contribute outputs
with quality less than $\at$, we can now conclude that 
$c_{\M}(1-F(\at))$ is equal to the expected
prize earned by a contestant with quality 
$q \le \at$ when the opponents play according to $\strat$.
If this were less than $c$, then it could not
be the case that $\strat(\at,c)=1$ in equilibrium.
On the other hand, if $c_{\M}(1-F(\at))$ were
greater than $c$, then the expected prize earned
by any participating agent with quality $q \leq \at$
is greater than $c$ and consequently all such agents
should participate in equilibrium, contradicting
the fact that $\strat(q,c)=0$ for $q < \at$. 
Since each of the inequalities $c_{\M}(1-F(\at))<c$
and $c_{\M}(1-F(\at))>c$ leads to a contradiction,
it must be the case that $c_{\M}(1-F(\at))=c$.
\endproof

Continuing with deferred proofs from \S \ref{s-homo}, let us
restate and prove Proposition~\ref{p-compstat}.

\newtheorem*{p-compstat}{Proposition \ref{p-compstat}}
\begin{p-compstat} 
Fix a total prize, $V$, and number of potential contestants, $n$.
Let $\M^j$ denote a simple contest that awards equals prizes $V/j$
to the top $j$ ranks. 
\begin{enumerate}
\item \label{i-compstat1}
There is a decreasing sequence
$V = c_1 > c_2 > \cdots  > c_n > c_{n+1}=0$, such
that for any participation cost $c < V$, the simple contest
$\M^j$ is optimal if and only if $c_{j+1} \leq c \leq c_j$.
\item \label{i-compstat2}
For $1 < j \leq n$, the equation $c_{\M^j}(p) = c_{\M^{j-1}}(p)$
has a unique solution $p_j$ in the interval $(0,1)$, and
$c_j = c_{\M^j}(p_j) = c_{\M^{j-1}}(p_j)$.
\end{enumerate}
\end{p-compstat}
\begin{proof}
Let us compare the simple contests $\M^j$
and $\M^{j-1}$, for $1 < j \leq n$. For a given
participation cost, $c$, let $q^j$ and $q^{j-1}$ 
denote the equilibrium thresholds of 
$\M^j$ and $\M^{j-1}$, respectively.
The probability that an individual agent
participates in the equilibrium of $\M^j$
(resp.\ $\M^{j-1}$) will be denoted by  $p^j = 1-F(q^j)$,
resp.\ $p^{j-1} = 1-F(q^{j-1})$.
From Proposition~\ref{p-eq-homo} we know that 
$c = c_{\M^j}(p^j) = c_{\M^{j-1}}(p^{j-1}),$ which 
implies
\BEQ \label{e-cs2-1}
0 = c_{\M^j}(p^j) - c_{\M^{j-1}}(p^{j-1}) =
[c_{\M^j}(p^j) - c_{\M^{j-1}}(p^j)] +
[c_{\M^{j-1}}(p^j) - c_{\M^{j-1}}(p^{j-1})].
\EEQ
Combining this equation with the fact that
the function $c_{\M^{j-1}}$ is strictly decreasing
(Lemma~\ref{l-qc1to1})
% so the second term on the
% right side of~\eqref{e-cs2-1} has the opposite
% sign from 
we obtain the following chain of equivalences:
\BEQ \label{e-cs2-2}
p^j > p^{j-1} \Longleftrightarrow
c_{\M^{j-1}}(p^j) < c_{\M^{j-1}}(p^{j-1}) \Longleftrightarrow
c_{\M^j}(p^j) - c_{\M^{j-1}}(p^j) > 0.
\EEQ
Thus, the question of which mechanism, $\M^j$ or $\M^{j-1}$, 
has a higher rate of participation in equilibrium
reduces to evaluating whether $c_{\M^j} - c_{\M^{j-1}}$ is 
positive or negative at $p^j$.
To understand the sign of the function 
$c_{\M^j} - c_{\M^{j-1}}$ on the interval $[0,1]$,
the following calculation is crucial.
\begin{align*}
c_{\M^j(p)} - c_{\M^{j-1}(p)} &=
\tfrac{V}{j} \sum_{k=1}^j \tbinom{n-1}{k-1} p^{k-1} (1-p)^{n-k} -
\tfrac{V}{j-1} \sum_{k=1}^{j-1} \tbinom{n-1}{k-1} p^{k-1} (1-p)^{n-k} \\
& =
\tfrac{V}{j} \tbinom{n-1}{j-1} p^{j-1} (1-p)^{n-j} +
\sum_{k=1}^{j-1} \left( \tfrac{V}{j} - \tfrac{V}{j-1} \right)
\tbinom{n-1}{k-1} p^{k-1} (1-p)^{n-k} \\
& = 
\tfrac{V}{j} (1-p)^{n-1} \tbinom{n-1}{j-1} \left( \tfrac{p}{1-p} \right)^{j-1}
- \tfrac{V}{j(j-1)} (1-p)^{n-1} \sum_{k=1}^{j-1} \tbinom{n-1}{k-1} 
\left( \tfrac{p}{1-p} \right)^{k-1} \\
& = 
\tfrac{V}{j(j-1)} (1-p)^{n-1} \left[
(j-1) \tbinom{n-1}{j-1} \left( \tfrac{p}{1-p} \right)^{j-1} -
\sum_{k=1}^{j-1} \tbinom{n-1}{k-1} 
\left( \tfrac{p}{1-p} \right)^{k-1} \right]
\end{align*}
Letting $x = \frac{p}{1-p}$ we have derived
\BEQ \label{e-cs2-3}
c_{\M^j(p)} - c_{\M^{j-1}(p)} =
\tfrac{V}{j(j-1)} (1-p)^{n-1} \left[
(j-1) \tbinom{n-1}{j-1} x^{j-1} - \sum_{k=1}^{j-1} 
\tbinom{n-1}{k-1} x^{k-1} \right]
% \tbinom{n-1}{j-2} x^{j-2} - \tbinom{n-1}{j-3} x^{j-3} - \cdots
% - \tbinom{n-1}{1} x - 1 \right] .
\EEQ
The polynomial $(j-1) \tbinom{n-1}{j-1} x^{j-1} -
\tbinom{n-1}{j-2} x^{j-2} - \tbinom{n-1}{j-3} x^{j-3} - \cdots
- \tbinom{n-1}{1} x - 1 $ appearing on the right
side of~\eqref{e-cs2-3} is negative at $x=0$, tends to
$+\infty$ as $x \to \infty$, and by Descartes' Rule of Signs,
it has a single positive real root, $r$. Accordingly, there is
a unique $p_j = \frac{r}{r+1}$ in the interval $(0,1)$ satisfying
the equation $c_{\M^{j}}(p_j) = c_{\M^{j-1}}(p_j)$, and
the quantity $c_{\M^j}(p) - c_{\M^{j-1}}(p)$ is positive if and
only if $p > p_j$. As we vary the participation cost, $c$,
the equilibrium participation probability of the simple 
contest $\M^j$, namely $p^j$, is a decreasing function of $c$. 
Therefore, letting $c_j = c_{\M^{j}}(p_j) = c_{\M^{j-1}}(p_j)$, we have the 
following chain of equivalences.
\BEQ \label{e-cs2-4}
p^j > p^{j-1} \Longleftrightarrow
c_{\M^j}(p^j) - c_{\M^{j-1}}(p^j) > 0 \Longleftrightarrow
p^j > p_j \Longleftrightarrow
c < c_j.
\EEQ
The proposition claims that the numbers $(c_j)$ form a
decreasing sequence. We now turn to the proof of this
fact. Recalling that the equation
$
  c_j = c_{\M^j}(p_j) = c_{\M^{j-1}}(p_j)
$
holds for all $j \geq 2$, we find that
\BEQ \label{e-cs2-5}
c_j - c_{j+1} = c_{\M^j}(p_j) - c_{\M^j}(p_{j+1}) .
\EEQ
The function $x(p) = \frac{p}{1-p}$ is an increasing
function of $p$, and $c_{\M^j}(p)$ is a decreasing
function of $p$. This justifies the second in the 
following pair of equivalences.
\BEQ \label{e-cs2-6}
c_j > c_{j+1} \; \stackrel{\eqref{e-cs2-5}}{\Longleftrightarrow} \; 
c_{\M^j}(p_j) > c_{\M^j}(p_{j+1})  \; \Longleftrightarrow \;
x(p_j) < x(p_{j+1}),
\EEQ
We are left with proving that the RHS of~\eqref{e-cs2-6}
is valid. For every $j \geq 2$ define the polynomial
\BEQ \label{e-cs2-7}
Q_j(x) = (j-1) \tbinom{n-1}{j-1} x^{j-1} - 
         \sum_{k=1}^{j-1} \tbinom{n-1}{k-1} x^{k-1}.
\EEQ
We know from~\eqref{e-cs2-3} and the ensuing paragraph
that the sign of $Q_j(x)$ is the same as the sign of
$x - x(p_j)$, so to prove that $x(p_j) < x(p_{j+1})$,
it suffices to show that $Q_j(x(p_{j+1})) > 0$. The easy
calculation
\BEQ \label{e-cs2-8}
Q_{j+1}(x) - Q_j(x) = 
j \tbinom{n-1}{j} x^j - j \tbinom{n-1}{j-1} x^{j-1} =
j \tbinom{n-1}{j} x^{j-1} \left( x - \tfrac{j}{n-j} \right),
\EEQ
combined with the fact that $Q_{j+1}(x(p_{j+1})) = 0$, 
reveals that $Q_j(x(p_{j+1})) > 0$ if and only if
$x(p_{j+1}) < \tfrac{j}{n-j}$. Recalling that 
$x(p_{j+1}) < x$ if and only if $Q_{j+1}(x) > 0$,
we are left with showing that $Q_{j+1} \left( \tfrac{j}{n-j} \right) > 0$.
The calculation is as follows.
\begin{align*}
Q_{j+1} \left( \tfrac{j}{n-j} \right) &= 
j \tbinom{n-1}{j} \left( \tfrac{j}{n-j} \right)^j -
\sum_{k=1}^{j} \tbinom{n-1}{k-1} \left( \tfrac{j}{n-j} \right)^{k-1} \\
&=
\sum_{k=1}^{j} \left[ \tbinom{n-1}{j} \left( \tfrac{j}{n-j} \right)^j
- \tbinom{n-1}{k-1} \left( \tfrac{j}{n-j} \right)^{k-1} \right] \\
&= \tbinom{n-1}{j} \left( \tfrac{j}{n-j} \right)^j 
\sum_{k=1}^j \left[ 1 - \tfrac{(n-1)!}{(k-1)! (n-k)!} \cdot
\tfrac{j! (n-1-j)!}{(n-1)!} \cdot \left( \tfrac{j}{n-j} \right)^{k-1-j} \right] 
\\ 
&= \tbinom{n-1}{j} \left( \tfrac{j}{n-j} \right)^j 
\sum_{k=1}^j \left[ 1 - \tfrac{j!}{(k-1)!} \cdot \tfrac{(n-1-j)!}{(n-k)!} \cdot
\left( \tfrac{j}{n-j} \right)^{k-1-j} \right] 
\\
&= \tbinom{n-1}{j} \left( \tfrac{j}{n-j} \right)^j 
\sum_{k=1}^j \left[ 1 - \left( \prod_{i=k}^{j} \tfrac{i}{n-i} \cdot 
\tfrac{n-j}{j} \right) \right] \\
&= \tbinom{n-1}{j} \left( \tfrac{j}{n-j} \right)^j 
\sum_{k=1}^j \left[ 1 - \prod_{i=k}^j \tfrac{in - ij}{jn - ij} \right] \\
&= \tbinom{n-1}{j} \left( \tfrac{j}{n-j} \right)^j
\sum_{k=1}^{j-1} \left[ 1 - \prod_{i=k}^j \tfrac{in-ij}{jn-ij} \right]
\end{align*}
Recall that we are assuming $j>1$, so the sum in the final line
is over a non-empty set of indices $k$.
Each term in the product on the final line is less than or equal to 1,
at least one of the terms is strictly less than 1, 
so the product is strictly less than 1.
This means that each summand on the final line
is strictly positive, so the entire quantity on the final line is
strictly positive, as desired. This completes the proof that
$Q_{j+1} \left( \tfrac{j}{n-j} \right) > 0$, and as we have seen
this implies that $c_j > c_{j+1}$ for $j > 1$. To complete the proof
that the entire sequence $c_1,c_2,\ldots,c_{n+1}$ is strictly decreasing,
we need only prove that $c_1 > c_2$. This inequality follows easily,
once we recall that $c_1=V$ while 
\[
c_2 = c_{\M^2}(p_2) = \tfrac{V}{2} \sum_{k=1}^2 \tbinom{n-1}{k-1} p_2^{k-1} (1-p_2)^{n-k} \leq \tfrac{V}{2} \sum_{k=1}^n \tbinom{n-1}{k-1} p_2^{k-1} (1-p_2)^{n-k} = 
\tfrac{V}{2}.
\]
It remains to show that $\M^j$ is an optimal simple contest
if and only if $c_{j+1} \leq c \leq c_j$. Recall that for
every $j>2$, 
$\M^j$ has a higher rate of participation than $\M^{j-1}$,
in equilibrium, if and only if $c < c_j$. This immediately
shows that $c_{j+1} \leq c \leq c_j$ is a necessary condition
for $\M^j$ to be optimal, as otherwise either $\M^{j+1}$ or
$\M^{j-1}$ is superior to $\M^j$. To prove that it is also a
sufficient condition, we will prove that $\M^j$ is strictly
optimal whenever $c_{j+1} < c < c_j$;
an easy continuity argument then confirms that $\M^j$ is weakly
optimal whenever $c_{j+1} \leq c \leq c_j$.

Suppose that $c_{j+1} < c < c_j$
and for every $k=1,\ldots,n$ let $p^k$ denote the equilibrium
participation probability for $\M^k$. For $2 \leq k \leq j$ we have
the string of inequalities 
$c \leq c_j < c_{j-1} < \cdots < c_{k+1} < c_k$, which
means that $c < c_k$ and hence that $p^{k-1} < p^k$. 
Stringing these inequalities together for all $k$ in the range
$2,\ldots,j$, we find that $p^1 < p^2 < \cdots < p^{j-1} < p^j$,
\ie, $\M^j$ has a higher rate of participation, in equilibrium,
than $\M^k$ for any $k<j$. The argument for $k>j$ is the same
with the inequality signs reversed. For $j < k \leq n$ we have
the string of inequalities $c \geq c_{j+1} > c_{j+2} \cdots > c_{k-1} > c_k$
which means that $c > c_k$ and hence that $p^{k-1} > p^k$. Stringing
these inequalities together for all $k$ in the range
$j+1,\ldots,n$, we find that $p^j > p^{j+1} > \cdots > p^{n-1} > p^n$,
\ie, $\M^j$ has a higher rate of participation, in equilibrium,
than $\M^k$ for any $k>j$.
\end{proof}

Proposition~\ref{p-quant} makes three assertions about 
optimal contests and their equilibria in the setting of
homogeneous costs. We now restate and prove
the first two parts of the proposition. The
following subsection presents the proof of the
third part.

\newtheorem*{p-quant-12}{Proposition \ref{p-quant} (parts 1-2)}
\begin{p-quant-12}
Fix $V,c$, and let $j^{*,n}$ denote the number of prizes
offered by the optimal contest with population size $n$.
Let $p^{*,n}$ denote the probability that an individual agent
participates in the equilibrium of that contest, and let
$\ld^{*,n} = n \!\cdot\! p^{*,n}$ denote the expected number
of participants in that equilibrium.
\begin{enumerate}
\item The winner-take-all contest is optimal if and only if
$\frac{V}{c} \leq \left( 1 + \frac{1}{n-1} \right)^{n-1}$. In particular,
for $\frac{V}{c} > e$ a winner-take-all contest is never optimal, whereas
for $\frac{V}{c} < e$ it is optimal whenever the population size, $n$,
is sufficiently large.
\item  As $n \to \infty$, the sequence
$\ld^{*,n}$ converges to the unique $\ld^* > 0$ satisfying
\BEQ \label{e-ldstar-appendix}
\max_{1 \leq j \leq V/c} \left\{ \tfrac{V}{j}  \sum_{k=0}^{j-1}
\tfrac{ e^{-\ld} \ld^k }{k!} \right\}  = c.
\EEQ
Excluding a countable set of values for $c$, the maximum
in~\eqref{e-ldstar} is achieved at a unique value $j^*$,
and the sequence $j^{*,n}$ converges to $j^*$ as $n \to \infty$.
\end{enumerate}
\end{p-quant-12}
\begin{proof}
Proposition~\ref{p-compstat} tells us that the winner-take-all
contest is optimal if and only if $c \geq c_2 = c_{\M^1}(p)$, 
where $p$ is the unique solution of the equation $c_{\M^1}(p) = c_{\M^2}(p)$.
We now calculate:
\begin{align*}
c_{\M^1}(p) - c_{\M^2}(p) & = 
V (1-p)^{n-1} - \frac{V}{2} \left[ (1-p)^{n-1} + (n-1) p (1-p)^{n-2} \right] \\
& =
\frac{V}{2} (1-p)^{n-2} \left[ 1 - np \right].
\end{align*}
Hence $p=\frac{1}{n}$ is the unique solution to $c_{\M^1}(p) = c_{\M^2}(p)$.
Now we find that 
\BEQ \label{e-quant-2}
c_2 = c_{\M^1}(p) = V (1-p)^{n-1} = V \left(1 - \tfrac{1}{n} \right)^{n-1}
\EEQ
so the requirement $c \geq c_2$ can be rewritten as
$c \geq V \left( \frac{n-1}{n} \right)^{n-1}$, which in
turn can be rewritten as 
$\frac{V}{c} \leq \left( 1 + \frac{1}{n-1} \right)^{n-1}$,
matching the criterion for optimality of winner-take-all 
claimed in part~\ref{i-quant-wta} of the proposition. The observation
that $\left( 1 + \frac{1}{n-1} \right)^{n-1}$ is always less
than $e$, and converges to $e$ as $n \to \infty$, supplies
the justification for the last sentence of part~\ref{i-quant-wta}.

Moving on to the second part of the proposition, let 
$p^{j,n}$ denote the equilibrium participation probability
in the simple contest $\M^j$ when the population size is $n$.
If $j > \vc = V/c$ then $V/j < c$, \ie, winning a prize
in $\M^j$ is not worth enough to cover the cost of participating.
Hence, $p^{j,n}=0$ if $j > \vc$. 
Using the fact that optimal
contests are simple and that a simple contest with $j > \vc$ prizes
has no participation, we find that
\BEQ \label{e-quant-3}
c = c^*(p^{*,n}) = \max_{1 \leq j \leq \vc} c^n_{\M^j}(p^{*,n}) ,
\EEQ
where $c^n_{\M^j}(\cdot)$ denotes the function $c_{\M}(\cdot)$
defined in Equation~\eqref{e-cm}, specialized to population
size $n$ and contest
$\M = \M^j$.
Under the change of variables
$\ld = \ld(p) = n \!\cdot\! p$,
the function $c^n_{\M^j}(p)$ is expressed by the formula
\BEQ \label{e-quant-4}
c^n_{\M^j}(p) = c^n_{\M^j} \left( \tfrac{\ld}{n} \right) =
\tfrac{V}{j}
\sum_{k=0}^{j-1} 
\tbinom{n-1}{k} \left( \tfrac{\ld}{n} \right)^{k}
\left( 1 - \tfrac{\ld}{n} \right)^{n-1-k} .
\EEQ
For any fixed $j$, as $n \to \infty$ the RHS of~\eqref{e-quant-4} 
converges uniformly as a function of $\ld \in [0,\vc]$. 
Indeed, for all fixed $k \in \naturals$,
\begin{align}
\label{e-quant-5.0}
\lim_{n \to \infty} \left( 1 - \tfrac{\ld}{n} \right)^{n-1-k} & =
\left[ \lim_{n \to \infty} \left( 1 - \tfrac{\ld}{n} \right)^n \right]
\cdot
\left[ \lim_{n \to \infty} \left( 1 - \tfrac{\ld}{n} \right)^{-1-k} \right] =
e^{-\ld} \\
\label{e-quant-5.1}
\lim_{n \to \infty} \tbinom{n-1}{k} \left( \tfrac{\ld}{n} \right)^{k} & =
\lim_{n \to \infty} \prod_{i=1}^{k} \left( \tfrac{n-i}{i} \cdot \tfrac{\ld}{n} \right) 
%  \\ \nonumber & 
= 
\tfrac{\ld^k}{k!}  \left(
\lim_{n \to \infty} \prod_{i=1}^{k} \tfrac{n-i}{n}  \right) =
\tfrac{\ld^k}{k!} \\
\label{e-quant-5.2}
\lim_{n \to \infty} c^n_{\M^j} \left( \tfrac{\ld}{n} \right) & =
\tfrac{V}{j} \sum_{k=0}^{j-1} \tfrac{e^{-\ld} \ld^k}{k!},
\end{align}
and the convergence on each line of~\eqref{e-quant-5.0}-\eqref{e-quant-5.2}
is uniform over $\ld \in [0,\vc]$. 

Let $c^\infty_j$ denote the function appearing on the right side
of~\eqref{e-quant-5.2}, i.e.\ $c^\infty_j(\ld) = \frac{V}{j}
\sum_{k=0}^{j-1} \tfrac{e^{-\ld} \ld^k}{k!} .$
This is a continuous and strictly
decreasing function of $\ld$,
for all $\ld \geq 0$. To verify this, we can simply take
the derivative with respect to $\ld$.
\begin{align}
\nonumber
\frac{d}{d \ld} \left[ \sum_{k=0}^{j-1} \tfrac{e^{-\ld} \ld^k}{k!} \right] 
 & =
- e^{-\ld} + \sum_{k=1}^{j-1} \left[ \tfrac{e^{-\ld} \ld^{k-1}}{(k-1)!} - 
\tfrac{e^{-\ld} \ld^k}{k!} \right] \\
\label{e-quant-6}
& =
e^{-\ld} \left[ \sum_{i=0}^{j-2} \tfrac{\ld^{i}}{i!} -
\sum_{k=0}^{j-1} \tfrac{\ld^{k}}{k!} \right] 
= - \tfrac{e^{-\ld} \ld^{j-1}}{(j-1)!}.
\end{align}
It is easy to see that $c^\infty_j(0) = \frac{V}{j}$ and
that $\lim_{\ld \to \infty} c^{\infty}_j(\ld) = 0$. Combined 
with the fact that $c^{\infty}_j$ is continuous and strictly
decreasing, this means it has a continuous and strictly
decreasing inverse function $\ld^j(c)$, defined for
$0 < c \leq \frac{V}{j}$.

Let $\ld^{j,n} = n \cdot p^{j,n}$. If $n(1) < n(2) < \cdots$ is any
infinite sequence of indices such that 
$\lim_{i \to \infty} \ld^{j,n(i)}$ exists,
then the uniform convergence in Equation~\eqref{e-quant-5.2}
combined with the equation $c^n_{\M^j}(p^{j,n}) = c$ which
is valid for all $j,n$, imply that
\BEQ
c = \lim_{i \to \infty} c^{n(i)}_{\M^j}(p^{j,n(i)}) 
  = \lim_{i \to \infty} c^{n(i)}_{\M^j} \left( \tfrac{\ld^{j,n(i)}}{n(i)} \right)
  = c^\infty_j \left( \lim_{i \to \infty} \ld^{j,n(i)} \right).
\EEQ
Recalling that the inverse function of $c^{\infty}_j$ is denoted by $\ld^j$, 
this means that $\lim_{i \to \infty} \ld^{j,n(i)} = \ld^j(c)$.

Recall now that $\ld^{j,n}$ can be interpreted as the expected
number of participants in the equilibrium of contest $\M^j$ when
the population size is $n$. This means that the expected total
participation cost is $\ld^{j,n} \cdot c$. Individual rationality
implies that $\ld^{j,n} \cdot c \leq V$, hence $\ld^{j,n} \leq V/c = \vc$.
Thus, every element of the infinite sequence $(\ld^{j,n})_{n=1}^{\infty}$ 
belongs to the interval $[0,\vc]$. Earlier we derived that
for every infinite subsequence that has a limit point, this limit
point must be equal to $\ld^j(c)$. It follows that the entire
sequence $\ld^{j,n}$ must converge to $\ld^j(c)$, as otherwise
there would be some $\varepsilon>0$ and an infinite subsequence
belonging 
to the set $[0,\ld^j(c) - \varepsilon] \cup [\ld^j(c)+\varepsilon,\vc]$.
That set is compact, so the infinite subsequence would have a limit
point in the set, contradicting our earlier derivation that every
subsequential limit point must be equal to $\ld^j(c)$.

Earlier we noted that $p^{j,n} = 0$ for $j > \vc$; hence
$p^{*,n} = \max_{1 \leq j \leq \vc} \{p^{j,n}\}$ and
$\ld^{*,n} = \max_{1 \leq j \leq \vc} \{\ld^{j,n}\}$.
This implies
\BEQ
\label{e-quant-7}
\lim_{n \to \infty} \ld^{*,n} = \lim_{n \to \infty} \left(
\max_{1 \leq j \leq \vc} \{ \ld^{j,n} \} \right) =
\max_{1 \leq j \leq \vc} \left\{ \lim_{n \to \infty} \ld^{j,n} \right\} =
\max_{1 \leq j \leq \vc} \{ \ld^j(c) \}.
\EEQ
We must show that the right side of~\eqref{e-quant-7} is
equal to the unique $\ld^* > 0$ satisfying Equation~\eqref{e-ldstar-appendix}.
Using notation we have introduced in this proof, that 
equation can be rewritten as
\BEQ \label{e-quant-8}
\max_{1 \leq j \leq \vc} \{c^\infty_j(\ld^*)\} = c.
\EEQ
The function 
$h(\ld) = \max_{1 \leq j \leq \vc} \{c^\infty_j(\ld)\}$,
being a pointwise maximum of continuous and strictly decreasing functions,
is also continuous and strictly decreasing. Thus, there is a unique
$\ld^* = \ld^*(c)$ satisfying~\eqref{e-quant-8}. 
Let $j^*$ be the value of $j$ that achieves the maximum in the
formula 
defining $h(\ld^*)$, so that $c = h(\ld^*) = c^{\infty}_{j^*}(\ld^*)$.
Applying the function $\ld^{j^*}$ to both sides of this equation 
yields $\ld^{j^*}(c) = \ld^*$, and therefore
$\ld^* \leq \max_{1 \leq j \leq \vc} \ld^j(c)$.
The inequality cannot be strict, because if 
$\ld^* < \ld^j(c)$ for some $j$, then we can apply the strictly
decreasing function $c^\infty_j$ to both sides and conclude
that $c^\infty_j(\ld^*) > c_j(\ld^j(c)) = c$, 
contradicting Equation~\eqref{e-quant-8}.

The proposition also asserts that
 with the exception of
countably many values of $c$, the maximum in~\eqref{e-ldstar-appendix}
is achieved by a unique value of $j$. To show this,
consider for any $j \neq \ell$ the equation
\BEQ \label{e-quant-7}
\tfrac{V}{j} \sum_{k=0}^{j-1} \frac{e^{-\ld} \ld^k}{k!} - 
\tfrac{V}{\ell} \sum_{k=0}^{\ell-1} \frac{e^{-\ld} \ld^k}{k!} = 0,
\EEQ
which must be satisfied by $\ld^*(c)$ whenever the maximum
in~\eqref{e-ldstar-appendix} is achieved at both $j$ and $\ell$.
For any two solutions $\ld_0,\ld_1$ 
of~\eqref{e-quant-7}, there must be a $\ld$ in the
open interval $(\ld_0,\ld_1)$ such that 
the derivative of the LHS of~\eqref{e-quant-7} 
equals zero at $\ld$. We know, from Equation~\eqref{e-quant-6},
that the derivative of the LHS of~\eqref{e-quant-7} is 
equal to $V e^{-\ld} \left( \frac{\ld^{\ell-1}}{\ell!} - \frac{\ld^{j-1}}{j!}
\right)$, which equates to zero only at $\ld = \left( \ell! / j!
\right)^{1/(\ell-j)}$. Thus, Equation~\eqref{e-quant-7} can have
at most two solutions for any fixed $j,\ell$. Taking the union
over all pairs $j,\ell$, we find that there are at most countably
many solutions to~\eqref{e-quant-7}. For any $c$ such that 
$\ld^*(c)$ does not belong to this countable set, the maximum
in~\eqref{e-ldstar-appendix} is attained at a unique
$j^* = j^*(c)$.

Finally we must prove that when $c$ lies outside this countable
set, the sequence $j^{*,n}$ converges to $j^*$ as $n \to \infty$.
Indeed, we have seen that each element of the sequence
$(j^{*,n})_{n=1}^{\infty}$ is an integer in the 
range $[1,\vc]$, so there must be at least one
$j \in [1,\vc]$ that occurs infinitely often in the sequence.
We will show that any such $j$ must equal $j^*$, from which
it follows immediately that $j^{*,n} \to j^*$ as $n \to \infty$.

Now suppose $n(1) < n(2) < \cdots$ is an infinite sequence
such that $j^{*,n(i)} = j$ for all $i \geq 1$. We have
\BEAS
\nonumber
\forall i \geq 1 \quad &
c^{n(i)}_{\M^j} \left( \tfrac{\ld^{*,n(i)}}{n} \right)  =
c^{n(i)}_{\M^j} \left( \tfrac{\ld^{j,n(i)}}{n} \right) =
c^{n(i)}_{\M^j} \left( p^{j,n(i)} \right) = c 
\EEAS
\begin{align}
\label{e-quant-9}
c^{\infty}_j ( \ld^* ) & =
c^{\infty}_j \left( \lim_{i \to \infty} \ld^{*,n(i)} \right) =
\lim_{i \to \infty} c^{n(i)}_{\M^j} \left( \tfrac{\ld^{*,n(i)}}{n} \right) = c
\end{align}
where the last line is justified by the uniform convergence
guarantee in Equation~\eqref{e-quant-5.2}.
We know that $c^{\infty}_{j^*}(\ld^*)=c$, and our
assumption on $c$ implies that for every $j' \neq j^*$,
$c^{\infty}_{j'}(\ld^*) < c$. Hence, Equation~\eqref{e-quant-9}
implies that $j = j^*$, which completes the proof of the
proposition.
\end{proof}

\subsection{Proof of Proposition~\ref{p-quant}, part \ref{i-quant-asymp}}

The proof of Proposition~\ref{p-quant}, part \ref{i-quant-asymp} 
depends on some rather sharp estimates of binomial coefficients.
To derive these estimates, we begin by deriving 
a non-asymptotic version of Stirling's approximation for $n!$.

\begin{lemma} \label{lem:factorial}
For any positive integer $n$, 
\begin{equation} \label{eq:log-factorial}
n \ln(n) - n + \tfrac12 \ln(n) + \tfrac23 < \ln(n!) < 
n \ln(n) - n + \tfrac12 \ln(n) + 1.
\end{equation}
\end{lemma}
\begin{proof}
For all $k$ and all $t \in (0,1)$ we have
\[
  \ln(k) + t \left(\ln(k+1) - \ln(k)\right) <
  \ln(k+t) <
  \ln(k) + \tfrac{t}{k},
\]
where the left inequality is derived from the strict concavity 
of the natural logarithm function, and the right inequality
is derived from strict concavity along with the fact that the
derivative of the natural logarithm at $k$ is $\frac{1}{k}$.
Integrating with respect to $t$ and applying the substitution
$x=k+t$, we find that
\begin{equation}
  \ln(k) + \tfrac12 \left( \ln(k+1) - \ln(k) \right) <
  \int_{k}^{k+1} \ln(x) \, dx  <
  \ln(k) + \tfrac{1}{2k}.
\end{equation}
Now, summing over $k=1,\ldots,n-1$, 
\begin{equation}
  \ln(n!) - \tfrac12 \ln(n) < 
  \int_{1}^{n} \ln(x) \, dx <
  \ln(n!) - \ln(n) + \frac12 \sum_{k=1}^{n-1} \frac{1}{k} <
  \ln(n!) - \tfrac12 \ln(n) + \tfrac{\gamma}{2},
\label{eq:factorial.2}
\end{equation}
where $\gamma = 0.577\ldots$ is the Euler-Mascheroni constant
and we have used the fact that $\sum_{k=1}^{n-1} \frac{1}{k} < \ln(n) + \gamma$.
Rearranging terms in~\eqref{eq:factorial.2} and
using the fact that $\int_1^n \ln(x) \, dx = 
n \ln(n) - n + 1$,
 we derive
\begin{equation}
  n \ln(n) - n + \tfrac12 \ln(n) + 1 - \tfrac{\gamma}{2} <
  \ln(n!) <
  n \ln(n) - n + \tfrac12 \ln(n) + 1
\end{equation}  
The lemma follows by observing that $\frac23 < 1 - \frac{\gamma}{2}$.
\end{proof}

\begin{lemma} \label{lem:bcoeff}
For any positive integers $n$ and $j \leq \frac{n}{2}$ and any $p \in (0,1)$, 
\begin{equation} \label{eq:bcoeff}
\binom{n}{j} p^j (1-p)^{n-j} > \frac14 \cdot \sqrt{\frac{1}{j}} \cdot 
\exp \left( - \frac{2(j-pn)^2}{pn} \right).
\end{equation}
Furthermore if $j > pn$ then 
\begin{equation} \label{eq:bcoeff.ub}
\binom{n}{j} p^j (1-p)^{n-j} < \exp \left( - \frac{(j-pn)^2-2}{2j} \right).
\end{equation}
\end{lemma}
\begin{proof}
Using Lemma~\ref{lem:factorial} and the formula
$\binom{n}{j} = \frac{n!}{j!(n-j)!}$, we obtain the bound
\begin{align}
\nonumber
\ln\binom{n}{j} &=
\ln(n!) - \ln(j!) - \ln((n-j)!) \\ 
\nonumber  & >
n \ln(n) - j \ln(j) - (n-j) \ln(n-j) + \tfrac12 \ln \left(
\tfrac{n}{j(n-j)} \right) - \tfrac43 \\
  & \geq
j \ln \left( \tfrac{n}{j} \right) +
(n-j) \ln \left( \tfrac{n}{n-j} \right) +
\tfrac12 \ln \left( \tfrac{1}{j} \right) - \tfrac43.
\label{eq:bcoeff.0}
\end{align}
Therefore the logarithm of the left side of~\eqref{eq:bcoeff} satisfies
\begin{equation} \label{eq:bcoeff.1}
\ln \binom{n}{j} + j \ln(p) + (n-j) \ln(1-p)  >
j \ln \left( \tfrac{pn}{j} \right) +
(n-j) \ln \left( \tfrac{(1-p)n}{n-j} \right) +
\tfrac12 \ln \left( \tfrac{1}{j} \right) - \tfrac43 
\end{equation}
Focusing attention on the first two terms on the
right side of~\eqref{eq:bcoeff.1}, we may perform the
following manipulation making use of the identity 
$\ln \left( \frac{1}{x} \right) = -\ln(x) \ge 1 - x$, which holds for all $x>0$.
\begin{align}
\nonumber
j \ln \left( \tfrac{pn}{j} \right) +
(n-j) \ln \left( \tfrac{(1-p)n}{n-j} \right) 
  &=
(j-pn) \ln \left( \tfrac{pn}{j} \right) +
(pn) \ln \left( \tfrac{pn}{j} \right) + \\
\nonumber  & \qquad \quad
(pn-j) \ln \left( \tfrac{(1-p)n}{n-j} \right) +
(n-pn) \ln \left( \tfrac{(1-p)n}{n-j} \right) \\
\nonumber  &\ge
(j-pn) \ln \left( \tfrac{pn(n-j)}{j(1-p)n)} \right) +
pn \left( 1 - \tfrac{j}{pn} \right) + 
(n-pn) \left(1 - \tfrac{n-j}{n-pn}\right) \\
\nonumber  &=
(j-pn) \ln \left( \tfrac{pn - pj}{j-pj} \right) +
pn - j + n - pn - (n-j) \\
\label{eq:bcoeff.2} &\ge
  (j-pn) \left( 1 - \tfrac{j-pj}{pn-pj} \right) =
  - \frac{(j-pn)^2}{p(n-j)}.
\end{align}
Combining~\eqref{eq:bcoeff.2} with~\eqref{eq:bcoeff.1},
we obtain
\begin{align} 
\nonumber
\ln \binom{n}{j} + j \ln(p) + (n-j) \ln(1-p) & > 
- \tfrac{(j-pn)^2}{p(n-j)} - \tfrac12 \ln(j) - \tfrac43 \\
 & >
- \tfrac{2(j-pn)^2}{pn} - \tfrac12 \ln(j) - \tfrac43,
\label{eq:bcoeff.3}
\end{align}
where the second inequality follows because $n-j \geq \frac{n}{2}$.
Exponentiating both sides of~\eqref{eq:bcoeff.3} and using
the fact that $e^{4/3} < 4$, we obtain the inequality~\eqref{eq:bcoeff}
stated in the lemma.

For the second part of the lemma, let $\ell = \lfloor pn \rfloor$.
The expression $\binom{n}{\ell} p^{\ell} (1-p)^{n-\ell}$ is 
the probability of observing $\ell$ heads when tossing a
bias-$p$ coin $n$ times. Using the trivial fact that this
probability is at most 1, we obtain the bound
\begin{align*}
\binom{n}{j} p^j (1-p)^{n-j} 
  & \le
\frac{\binom{n}{j} p^j (1-p)^{n-j}}{\binom{n}{\ell} p^{\ell} (1-p)^{n-\ell}} 
  =
\frac{\ell! (n-\ell)! p^j (1-p)^{n-j}}{j!(n-j)! p^{\ell} (1-p)^{n-\ell}} \\
  & =
\prod_{k=\ell+1}^{j} \frac{(n-k)p}{k(1-p)} 
   =
\prod_{k=\ell+1}^{j} \left( 1 - \frac{k-pn}{k-pk} \right) 
  <
\prod_{k=\ell+1}^{j} \left( 1 - \frac{k-pn}{j} \right) \\
  & <
\exp \left( - \frac{1}{j} \sum_{k=\ell+1}^{j} (k-pn) \right) 
  \le
\exp \left( - \frac{1}{2j} [(j-pn)^2 - 2] \right)
\end{align*}
which confirms~\eqref{eq:bcoeff.ub}.
%% For the second part of the lemma we use similar reasoning
%% to obtain an upper bound on the logarithm.
%% Using Lemma~\ref{lem:factorial}  we obtain the bound
%% \begin{align}
%% \nonumber
%% \ln\binom{n}{j} &=
%% \ln(n!) - \ln(j!) - \ln((n-j)!) \\ 
%% \nonumber  & <
%% n \ln(n) - j \ln(j) - (n-j) \ln(n-j) + \tfrac12 \ln \left(
%% \tfrac{n}{j(n-j)} \right) - \tfrac13 \\
%%   & =
%% j \ln \left( \tfrac{n}{j} \right) +
%% (n-j) \ln \left( \tfrac{n}{n-j} \right) +
%% \tfrac12 \ln \left( \tfrac{n}{j(n-j)} \right) - \tfrac13.
%% \label{eq:bcoeff.4}
%% \end{align}
%% Therefore the logarithm of the left side of~\eqref{eq:bcoeff} satisfies
%% \begin{equation} \label{eq:bcoeff.5}
%% \ln \binom{n}{j} + j \ln(p) + (n-j) \ln(1-p)  <
%% j \ln \left( \tfrac{pn}{j} \right) +
%% (n-j) \ln \left( \tfrac{(1-p)n}{n-j} \right) +
%% \tfrac12 \ln \left( \tfrac{n}{j(n-j)} \right) - \tfrac13 .
%% \end{equation}
%% Focusing attention on the first two terms on the right side
%% of~\eqref{eq:bcoeff.5}, we may use the identity $\ln(x) \le 1-x$
%% to derive
%% \begin{equation} \label{eq:bcoeff.6}
%% j \ln \left( \tfrac{pn}{j} \right) +
%% (n-j) \ln \left( \tfrac{(1-p)n}{n-j} \right) \le
%% j \left( 1 - \tfrac{pn}{j} \right) +
%% (n-j) \left( 1 - \tfrac{n-pn}{n-j} \right) \le
%% j - pn + n - j - (n - pn) = 0.
%% \end{equation}
%% Combining~\eqref{eq:bcoeff.6} with~\eqref{eq:bcoeff.5}
%% we obtain
%% \begin{equation} \label{eq:bcoeff.7}
%% \ln \binom{n}{j} + j \ln(p) + (n-j) \ln(1-p)  
%% <
%% \tfrac12 \ln \left( \tfrac{n}{j(n-j)} \right) - \tfrac13 \le
%% \tfrac12 \ln \left( \tfrac{2}{j} \right) - \tfrac13.
%% \end{equation}
 \end{proof}

\begin{lemma} \label{lem:avg}
Let $x_1,x_2,\ldots$ be any sequence of numbers
and for each integer $j > 0$ let $y_j = \frac1j \sum_{k=1}^{j} x_k$.
Then $y_{j+1} - y_j = \frac{1}{j+1}(x_{j+1} - y_j)$. In particular,
$y_{j+1} > y_j$ if and only if $x_{j+1} > y_j$.
\end{lemma}
\begin{proof}
The lemma follows from a simple calculation.
\begin{align*}
y_{j+1} - y_j & = \left( \frac{1}{j+1} \sum_{k=1}^{j+1} x_k \right)
   - \left( \frac{1}{j} \sum_{k=1}^{j} x_k \right) \\
  & = \frac{x_{j+1}}{j+1} - \left( \frac{1}{j} - \frac{1}{j+1} \right) 
        \sum_{k=1}^{j} x_k \\
  & = \frac{1}{j+1} \left( x_{j+1} - \frac{1}{j} \sum_{k=1}^{j} x_k \right) 
    = \frac{1}{j+1} \left( x_{j+1} - y_j \right).
\end{align*}
\end{proof}

\begin{proposition} \label{p-jstar}
For a given $n$ and $p$, if the value $j^* = j^*(n,p)$
defined by~\eqref{e-optw} satisfies $j^* \le n/2$ then
it satisfies
\[
j^* > p(n-1) + \frac12 \sqrt{ p(n-1) \ln^+ \left(
  \tfrac{p(n-1)}{16} \right) },
\]
where the notation $\ln^+(x)$ is defined to be equivalent
to $\max\{0,\ln(x)\}$.
\end{proposition}
\proof
We will be applying Lemma~\ref{lem:avg} to the
sequence
$x_j = \binom{n-1}{j-1} p^{j-1} (1-p)^{n-j}$
and its associated sequence of averages,
$y_j = \frac{1}{j} \sum_{k=1}^{j} x_k$.
Note that $j^* = j^*(n,p)$ defined % in Section~\ref{sec:compstat}
by~\eqref{e-optw}
is equal to $\argmax_j \{ y_j \}$.
The sequence $x_k$ is increasing for $k < pn$,
as can be seen from the calculation
\begin{align*}
  \frac{x_{k+1}}{x_{k}} & = 
  \frac{\binom{n-1}{k} p^{k} (1-p)^{n-k-1}}
       {\binom{n-1}{k-1} p^{k-1} (1-p)^{n-k}} \\
  & =
  \left( \frac{p}{1-p} \right) \left( \frac{(n-1)!}{k!(n-1-k)!} \right)
 \left( \frac{(k-1)! (n-k)!}{(n-1)!} \right) \\
  & =
  \frac{p}{1-p} \cdot \frac{n-k}{k} = \frac{pn-pk}{k-pk}.
\end{align*}
Thus, when $j < pn$, we have that $x_{j+1}$ is greater than
each of $x_1,\ldots,x_j$ hence also greater than their 
average, $y_j$. By Lemma~\ref{lem:avg} this means that
$j^* \geq pn$. 

Now, by another application of Lemma~\ref{lem:avg},
the fact that $y_{j^*+1} \leq y_{j^*}$ implies that
$x_{j^*+1} \leq y_{j^*}$. Noting that $y_{j^*} \leq \frac{1}{j^*}$,
and recalling our assumption that $j^* \leq n/2$, we may now
apply the lower bound for $x_{j^*+1}$ supplied by
Lemma~\ref{lem:bcoeff} to derive
\begin{align}
\nonumber
\frac{1}{j^*} \geq \binom{n-1}{j^*} p^{j^*} (1-p)^{n-1-j^*} & >
\frac{1}{4} \cdot \sqrt{\frac{1}{j^*}} \cdot \exp \left(
- \frac{2 (j^*-p(n-1))^2}{p(n-1)} \right) \\
\nonumber
\ln(j^*) & < \ln(4) + \tfrac12 \ln(j^*) + \frac{2 (j^*-p(n-1))^2}{p(n-1)} \\
\nonumber
\left[ \tfrac12 \ln(j^*) - \ln(4) \right] p (n-1) & < 2 (j^* - p(n-1))^2 \\
\nonumber
j^* & > p(n-1) + \sqrt{ \max\{0,\left[ \tfrac14 \ln(j^*) - \ln(2) \right]\} 
p (n-1)} \\
\label{eq:jstar.lb}
j^*    & > p(n-1) + \tfrac12 \sqrt{ \ln^+(p(n-1)/16) p (n-1) }.
\end{align}
\endproof

\begin{lemma} \label{l-btail}
Suppose we are given an integer $n$, a real number $p \in (0,1)$, and 
an integer $j$ such that $pn < j < n/2$. If
$\frac2j \leq \binom{n}{j} p^j (1-p)^{n-j} \leq \frac1j$, then
there exist absolute constants $C_0, C_1$ such that
\begin{equation} \label{e-btail}
\sqrt{\frac{C_0}{j \ln(j)}} \leq
\sum_{k=j}^{n} p^k (1-p)^{n-k} \leq
\sqrt{\frac{C_1}{j \ln(j)}}.
\end{equation}
\end{lemma}
\begin{proof}
Recall from the proof of Proposition~\ref{p-jstar} that if
$x_k$ is defined by $x_k = \binom{n}{k} p^k (1-p)^{n-k}$ then
\begin{align} 
\label{e-btail.0}
\frac{x_{k+1}}{x_k} = \frac{pn-pk}{k-pk} &= 1 - \frac{k-pn}{k-pk} 
\end{align}
The identity $2x \geq -\ln(1-x)$ is valid for
$ 0 \leq x \leq \frac34$. So when $k-pn < \frac34 (k-pk)$
we may apply this identity to conclude
\begin{align}
\label{e-btail.1}
2 \left( \tfrac{k-pn}{k-pk} \right) &\geq 
\ln x_{k} - \ln x_{k+1} 
% &\leq
% \tfrac{k-pn}{k-pk} .
\end{align}
In particular, for $\ell = \left\lceil j + \sqrt{ \frac{j}{\ln(j)} } 
\, \right\rceil$
we may sum the first inequality in~\eqref{e-btail.1}
over $k=j,j+1,\ldots,\ell-1$ to obtain
\begin{align}
\label{e-btail.2}
\ln x_{j} - \ln x_{\ell} & \leq
2 \sum_{k=j}^{\ell-1} \tfrac{k-pn}{k-pk} \\
& \leq
\frac{2}{j-pj} \sum_{k=j}^{\ell-1} (k-pn) \\
& = 
\frac{2 (j-pn) (\ell-j)}{j-pj} + 
\frac{2}{j-pj}
\sum_{k=j}^{\ell-1} (k-j) \\
& <
\frac{2 (j-pn) (\ell-j)}{j-pj} + \frac{(\ell-j)^2}{j-pj}.
\end{align}
Our hypothesis that $x_j = \Theta(1/j)$ implies
that $j - pn = \Theta(\sqrt{j \ln(j)})$ as was
shown earlier. Hence 
$$
(j-pn)(\ell-j) = \Theta \left(
  \sqrt{j \ln(j)} \cdot \sqrt{\frac{j}{\ln(j)}}
\right) = \Theta(j)
$$
while $(\ell-j)^2 = \frac{j}{\ln(j)}$. Combining
these bounds, and noting that the hypotheses in the
lemma statement imply $p \leq 1/2$ and hence $\frac{2}{1-p} = \Theta(1)$,
we find that
$$
\ln x_j - \ln x_{\ell} = \Theta(1),
$$
\ie, $x_{\ell} = \Theta(x_j)$.
The sequence $x_j,x_{j+1},\ldots,x_{\ell}$ is monotonically
decreasing, and we have derived that every term in this
sequence is $\Theta(x_j) = \Theta(1/j)$, so we have
shown that
\begin{equation} \label{e-btail.3}
x_j + \cdots + x_{\ell} = \Theta \left( \tfrac{\ell-j}{j} \right) 
= \Theta \left( \sqrt{\frac{1}{j \ln(j)}} \right). 
\end{equation}
The binomial tail $\sum_{k>j} \binom{n}{k} p^k (1-p)^{n-k}$
is equal to $x_j + \cdots + x_n$, so equation~\eqref{e-btail.3}
implies the existence of an absolute constant $C_0$ such that
\begin{equation} \label{e-btail.4}
\sqrt{ \frac{C_0}{j \ln(j)} } \leq \sum_{k>j} \binom{n}{k} p^k (1-p)^{n-k},
\end{equation}
establishing the first half of the lemma.

For the second half of the lemma, \ie, the upper bound on the binomial tail, 
we return to~\eqref{e-btail.0} and apply the
identity $-\ln(1-x) \geq x$, which is valid for
all $x \geq 0$. Hence for $k \geq j$,
\begin{align} \label{e-btail.4}
\ln x_k - \ln x_{k+1} &\geq \frac{k-pn}{k-pk} \geq \frac{k-pn}{2(j-pj)}.
\end{align}
For any $\ell>j$ we may sum over $k=j,\ldots,\ell-1$ to conclude
\begin{align*} 
\ln x_j - \ln x_{\ell} & \geq \frac{1}{j-pj} \sum_{k=j}^{\ell-1} (k-pn) \\
  & \geq
  \frac{1}{j-pj} \sum_{k=j}^{\ell-1} (j-pn) \\
  & = \frac{(\ell-j)(j-pn)}{j-pj} \\
\ln x_{\ell} & \leq \ln x_j - \left( \frac{j-pn}{j-pj} \right) (\ell-j).
\end{align*}
Let $s = \frac{j-pn}{j-pj}$. We have
\begin{align} \nonumber
\ln x_{\ell} & \leq \ln x_j - s (\ell-j) \\
\nonumber
x_{\ell} & \leq x_j \exp \left( - s (\ell-j) \right) \\
\nonumber
\sum_{\ell \geq j} x_{\ell} & \leq x_j \sum_{\ell=j}^{\infty} 
\exp \left( -s (\ell-j) \right) \\
\nonumber
 & < x_j \left[ 1 + \int_{0}^{\infty} e^{-sx} \, dx \right] \\
\label{e-btail.6}
 & = x_j (1 + \tfrac{1}{s}).
\end{align}
To finish, we estimate the terms appearing on the right
side of~\eqref{e-btail.6}. Recall $x_j = \Theta(1/j)$.
Our estimate $j-pn = \Theta(\sqrt{j \ln(j)})$ combined
with the fact that $1-p \geq \frac12$ implies that
$s = \Theta(\sqrt{\ln(j)/j})$. Therefore,
$ 1 + \frac1s = \Theta(\sqrt{j/\ln(j)})$ and
\[
\sum_{k=j}^{n} \binom{n}{k} p^k (1-p)^{n-k} = \sum_{\ell=j}^n x_{\ell}
  < x_j (1 + \tfrac{1}{s}) = \Theta \left( \frac1j \cdot \sqrt{\frac{j}{\ln(j)}}
\right)  = \Theta \left( \sqrt{\frac{1}{j \ln(j)}} \right),
\]
which shows that the upper bound in~\eqref{e-btail} holds
for a suitable choice of absolute constant $C_1$.
\end{proof}

\begin{lemma} \label{lem:ld.lb}
For a given total prize $V$ and participation cost $c$, 
let $\vc = V/c$
and $j = \lfloor \vc - \sqrt{\vc} \rfloor$. Consider a simple contest
that splits the prize into $j$ equal shares of size $V/j$.
In the equilibrium of this contest, the ex-ante participation
rate is greater than $\frac{1}{n-1} (\vc - \sqrt{5 \vc \ln(\vc)})$.
\end{lemma}
\begin{proof}
Let $p = \frac{1}{n-1} (\vc - \sqrt{5 \vc \ln(\vc)})$.
In Section~\ref{s-homo} we saw that to prove the lemma, it
is sufficient to prove that 
\begin{align}
\nonumber
\frac{V}{j} \sum_{k=1}^{j} \binom{n-1}{k-1} p^{k-1} (1-p)^{n-k} &\geq c \\
\nonumber
\sum_{k=1}^{j} \binom{n-1}{k-1} p^{k-1} (1-p)^{n-k} &\geq \frac{cj}{V} =
\nonumber
\frac{j}{\vc} \leq 1 - \sqrt{\frac{1}{\vc}} \\
\label{eq:ld.lb}
\sum_{k=j+1}^{n} \binom{n-1}{k-1} p^{k-1} (1-p)^{n-k} &\leq \sqrt{\frac{1}{\vc}}.
\end{align}
The left side is the probability of seeing at least $j$ heads
when observing $n-1$ tosses of a coin with bias $p$. Letting
$\ld = p(n-1)$ denote the expected number of heads and
$\delta = \frac{j}{\ld} - 1$, the Chernoff bound (e.g.~\cite{MR95})
tells us that
\[
\sum_{k=j+1}^{n} \binom{n-1}{k-1} p^{k-1} (1-p)^{n-k} \leq
e^{-\delta^2 \ld / 3}.
\]
To obtain an upper bound on $e^{-\delta^2 \ld/3}$, we 
derive a lower bound on $\delta^2 \ld$ as follows.
\begin{align*}
\delta \ld & = j - \ld \geq \vc - \sqrt{\vc} - 1 - 
  (\vc - \sqrt{5 \vc \ln(\vc)}) = \sqrt{5 \vc \ln(\vc)} - \sqrt{\vc} - 1 \\
\delta^2 \ld &= \frac{(\delta \ld)^2}{\ld} > \frac{(\delta \ld)^2}{\vc} \\
  &= \left( \frac{\delta \ld}{\sqrt{\vc}} \right)^2 \\
  &\geq \left( \sqrt{5 \ln(\vc)} - 1 - \vc^{-1/2} \right)^2 
%  &=  5 \ln(\vc) - 2 (1 + \vc^{-1/2}) \sqrt{5 \ln(\vc)} + 1 + 2 \vc^{-1/2} + \vc^{-1}.
\end{align*}
The expression on the last line is clearly asymptotic to $5 \ln(\vc)$ 
as $\vc \to \infty$, so for all sufficiently large $\vc$ it exceeds
$\frac32 \ln(\vc)$. In fact, one can verify numerically that
$\left( \sqrt{5 \ln(\vc)} - 1 - \vc^{-1/2} \right)^2 > \frac32 \ln(\vc)$
for all $\vc \geq 7$. So, to prove the lemma, we divide into two cases.
If $\vc < 7$ then $\vc - \sqrt{5 \vc \ln(\vc)} < 0$, so the lemma's
conclusion holds vacuously. If $\vc \geq 7$ then
$\delta^2 \ld > \frac32 \ln(\vc)$, so
\[
\sum_{k=j+1}^{n} \binom{n-1}{k-1} p^{k-1} (1-p)^{n-k} \leq
e^{-\delta^2 \ld / 3} < e^{- \ln(\vc)/2} = \sqrt{\frac{1}{\vc}},
\]
which establishes~\eqref{eq:ld.lb} and thus completes the proof.
\end{proof}

Using the foregoing lemmas, we now complete the proof
of Proposition~\ref{p-quant}, part~\ref{i-quant-asymp},
which we restate here for convenience.

\newtheorem*{p-quant-3}{Proposition \ref{p-quant}, part \ref{i-quant-asymp}}
\begin{p-quant-3}
Fix $V,c$, and let $j^{*,n}$ denote the number of prizes
offered by the optimal contest with population size $n$.
Let $p^{*,n}$ denote the probability that an individual agent
participates in the equilibrium of that contest, and let
$\ld^{*,n} = p^{*,n} \!\cdot\! n$ denote the expected number
of participants in that equilibrium.
Letting $\vc = V/c$, for any $n > 2 \vc$ 
the values of $\ld^{*,n}$ and $j^{*,n}$
satisfy 
\BEQ \label{e-asymptotic-append}
\ld^{*,n} = \vc - \Theta(\sqrt{\vc \log \vc}), \qquad
j^{*,n} = \vc - \Theta(\sqrt{\vc / \log \vc}).
\EEQ
\end{p-quant-3}
\begin{proof}
We begin by deriving the upper and lower bounds on $\ld=\ld^{*,n}$.
To obtain a lower bound, we use Lemma~\ref{lem:ld.lb}, 
% from Appendix~\ref{s-compstat-proofs}, 
which says
that a simple contest with $j=\lfloor \vc - \sqrt{\vc} \rfloor$
achieves participation rate greater than
$\frac{1}{n-1} ( \vc - \sqrt{5 \vc \ln(\vc)} )$.
The optimal participation rate $p$ can be no less than
this quantity, and consequently 
\begin{equation} \label{e-ld-lb}
\ld = pn > p(n-1) \geq \vc - \sqrt{5 \vc \ln(\vc)}.
\end{equation}
Before deriving an upper bound on $\ld$ we make some 
observations which lead to a crude lower bound on 
$j^* = j^{*,n}$.
Observe that a winner in the optimal contest
receives $V/j^*$ and has participation cost $c$, so individual rationality
guarantees that $V/j^* \geq c$, which we rewrite as
\begin{equation} \label{e-jstar-2}
j^* \leq V/c = \vc.
\end{equation}
By assumption, $\vc \leq n/2$, so we may apply Proposition~\ref{p-jstar}
to derive
%% recalling from \S \ref{s-homo}
%% that $j^* = j^{*,n}$, the number of prizes
%% in the optimal contest, satisfies
%% \BEQ \label{e-optw-append}
%% j^* = \argmax_{1 \le j \le n} 
%% \left\{ \tfrac{1}{j}\sum_{k=1}^j \tbinom{n-1}{k-1} p^{k-1}(1-p)^{n-k} 
%% \right\}.
%% \EEQ
%% Observe that a winner in the optimal contest
%% receives $V/j^*$ and has participation cost $c$, so individual rationality
%% guarantees that $V/j^* \geq c$, which we rewrite as
%% \begin{equation} \label{e-jstar-2}
%% j^* \leq V/c = \vc.
%% \end{equation}
%% A lower bound on $j^*$ can be obtained using~\eqref{e-optw-append}
%% as a starting point. Appendix~\ref{s-compstat-proofs} 
%% contains the calculations, which lead to the result
\begin{equation} \label{e-jstar-lb}
j^* > \ld + \tfrac12 \sqrt{\ld \ln^+(\ld/16)},
\end{equation}
where $\ln^+(x)$ is shorthand for $\max \{\ln(x),0\}$.
Combining~\eqref{e-jstar-lb} with~\eqref{e-jstar-2}
we obtain
\begin{equation} \label{e-jstar-3}
\vc > \ld + \tfrac12 \sqrt{\ld \ln^+(\ld/16)}.
\end{equation}
When $\vc \geq 100$, the lower bound~\eqref{e-ld-lb} 
implies that $\ld \geq \vc/2$.
Substituting this lower bound into~\eqref{e-jstar-3} implies
\begin{equation} \label{e-jstar-6}
\ld  < \begin{cases}
\vc - \tfrac12 \sqrt{\tfrac{\vc}{2} \ln^+ \left( \tfrac{\vc}{32} \right)} &
\mbox{if $\vc \geq 100$} \\
100 & \mbox{otherwise}
\end{cases}
\end{equation}
Combining~\eqref{e-ld-lb} with~\eqref{e-jstar-6} 
%% and simplifying,
we obtain
%% \begin{equation} \label{e-ld-lbub}
%% \vc - \sqrt{5 \vc \ln(\vc)} \le
%% \ld < 
%% \vc - \sqrt{\tfrac{\vc}{8} \ln^+ \left( \tfrac{\vc}{32} \right)} +100.
%% \end{equation}
% In particular, 
that $\ld = \vc - 
\Theta \left( \sqrt{\vc \ln(\vc)} \right)$
as $\vc \to \infty$.

As for estimating the number of prizes 
in the optimal contest, $j^*$,
we begin by recalling from Proposition~\ref{p-eq-homo}
that the participation rate $p$ in the equilibrium of the
optimal contest $\M$ satisfies $c_{\M}(p) = c$. When $\M$
is a simple contest that splits the prize into $j^*$ equal
shares of size $j^*$, we have 
$$c = c_{\M}(p) = \frac{V}{j^*} 
\sum_{k=1}^{j^*} \binom{n-1}{k-1} p^{k-1} (1-p)^{n-k}$$
and therefore
\begin{equation} \label{e-jstar-btail}
j^* = \frac{V}{c} \sum_{k=1}^{j^*} \binom{n-1}{k-1} p^{k-1} (1-p)^{n-k}
    = \vc \left[ 1 - \sum_{k>j^*} \binom{n-1}{k-1} p^{k-1} (1-p)^{n-k} \right].
\end{equation}
This shows that obtaining an asymptotic estimate of 
$\vc - j^*$ as a function of $\vc$ (as $\vc \to \infty$) 
is equivalent to estimating the binomial tail
$\sum_{k>j^*} \binom{n-1}{k-1} p^{k-1} (1-p)^{n-k}$
in terms of $\vc$. 
%% Unfortunately crude binomial tail
%% estimates such as those derived from Chernoff bounds
%% are insufficient for our purposes. A sharper tail bound
%% that is suitable for our purposes is
%% \begin{equation} \label{e-btail-theta}
%% \sum_{k>j^*} \binom{n-1}{k-1} p^{k-1} (1-p)^{n-k} =
%% \Theta \left( \sqrt{\frac{1}{j^* \ln(j^*)}} \right).
%% \end{equation}
%% The proof of this bound is given in the
%% Appendix, Lemma~\ref{l-btail}. The following
%% proposition concludes our analysis of $j^*$
%% as a function of $\vc = V/c$.
Lemma~\ref{l-btail} establishes that 
\BEQ
\sum_{k>j^*} \binom{n-1}{k-1} p^{k-1} (1-p)^{n-k} = 
\Theta \left( \sqrt{\tfrac{1}{j^* \ln(j^*)}} \right)
\EEQ
and substituting this estimate in~\eqref{e-jstar-btail} yields
\BEQ \label{e-jstar-penult}
j^* = \vc - \Theta \left( \sqrt{\tfrac{\vc}{j^* \ln(j^*)}} \right).
\EEQ
%% Let $\vc = V/c$ and assume $\vc \leq n/2$. 
%% The number of prizes in the optimal contest, $j^*$,
%% satisfies $j^* = \vc - \Theta(\sqrt{\vc/\ln(\vc)})$.
%% \end{proposition}
%% \begin{proof}
We have already shown above that for sufficiently large
$\vc$, $\ld = \Theta(\vc)$ and $j^* = \Theta(\ld) = \Theta(\vc)$.
Thus, $j^* \ln(j^*) = \Theta( \vc \ln(\vc) )$, and we can
substitute this estimate into~\eqref{e-jstar-penult} to derive
$j^* = \vc - \Theta(\sqrt{\vc/\ln(\vc)})$.
\end{proof}

\section{Equilibrium Existence and Uniqueness}
\label{s-exist-proofs}

In this section we restate and prove three lemmas related to
existence and uniqueness of symmetric equilibria in rank-order
contests. Throughout this section we assume that the distribution
$F(q,c)$ is absolutely continuous with respect to Lebesgue measure.

\begin{lemma} 
Any rank-order contest $\M$ has a symmetric mixed-strategy
equilibrium.
\end{lemma}
\begin{proof}
The lemma essentially follows from \citeauthor{milgrom85}'s
\citeyear{milgrom85} mixed equilibrium existence theorem for
games of incomplete information. Their theorem requires two
hypotheses: equicontinuous payoffs (R1) and absolutely continuous
information (R2). The game induced by contest $\M$ is easily
seen to satisfy both of these hypotheses. R1 is satisfied because
each player has only two actions (Proposition 1(a) of~\cite{milgrom85})
and R2 is satisfied because agents' types are independent
(Proposition 3(b) of~\cite{milgrom85}). The conclusion of
their theorem asserts that a mixed-strategy equilibrium exists
and does not specify existence of a symmetric equilibrium for
symmetric games such as ours. However, the existence of
a symmetric equilibrium is established by applying Glicksberg's
Fixed Point Theorem~\cite{glicksberg} in exactly the same
way that it is applied in the proof of Theorem 1 of~\cite{milgrom85},
except that the fixed point theorem is applied to the 
point-to-convex-set mapping that maps every mixed strategy $\mu$
(of a single agent) to that agent's best response set when
responding to the symmetric strategy profile in which all
agents play $x$.
\end{proof}

\begin{lemma} 
If $\mu$ is any symmetric
mixed strategy equilibrium of a rank-order contest $\M$ 
then there exists a symmetric
pure strategy equilibrium $\strat$ that is ``almost
equal'' to $\mu$ in the following sense: there exists a
measure-zero set of types, $T$, such that for all $(q,c) \not\in T$,
when $\strat'$ is randomly sampled from $\mu$, the event
$\strat'(q,c) \ne \strat(q,c)$ has probability zero.
\end{lemma}
\begin{proof}
Define $F_\mu$ to be the cumulative distribution function
of the random variable $q \cdot \strat'(q,c)$ when $(q,c)$
is drawn from $F$ and $\strat'$ is drawn from $\mu$. As in
the proof of Lemma~\ref{l-se}, an agent of type $(q,c)$
facing $n-1$ opponents all of whom are playing mixed
strategy $\mu$ receives an expected prize equal to
$c_{\M}(1-F_\mu(q))$ if she participates. If 
$c > c_{\M}(1-F_\mu(q))$ then the agent strictly
prefers non-participation to participation, hence
$\strat'(q,c)=0$ for almost every $\strat'$ in $\mu$.
If $c < c_{\M}(1-F_\mu(q))$ then the agent strictly
prefers participation to non-participation, hence
$\strat'(q,c)=1$ for almost every $\strat'$ in $\mu$.
This verifies that the lemma's conclusion holds with
$$
\strat(q,c) = \begin{cases}
  1 & \mbox{if } c < c_{\M}(1-F_\mu(q)) \\
  0 & \mbox{if } c \ge c_{\M}(1-F_\mu(q))
\end{cases}
$$
and $T = \{(q,c_{\M}(1-F_\mu(q))\}$. This set $T$ has measure
zero because it intersects each vertical line $\{q\} \times \reals$
in a single point.
\end{proof}

\begin{lemma} 
If $\M$ is any rank-order contest and 
$\strat, \strat'$ are two of its symmetric pure strategy equilibria
then $\{(q,c) \mid \strat(q,c) \ne \strat'(q,c)\}$ has measure zero.
\end{lemma}
\begin{proof}
Since $\strat,\strat'$ are both equilibria, they are also both 
sub-equilibria. Let $F_{\strat}, F_{\strat'}$ denote the cumulative
distribution functions of $q \cdot \strat(q,c)$ and
$q \cdot \strat'(q,c)$, respectively. Lemma~\ref{l-se} 
applied to the sub-equilibrium $\strat$ implies
$F_{\strat}(x) \ge F_{\strat'}(x)$ for all $x$. 
Applying the same lemma to the sub-equilibrium $\strat'$
yields the reverse inequality. Hence $F_{\strat} = F_{\strat'}$.
Reasoning now as in the proof of Lemma~\ref{l-pure},
we may conclude that $\strat(q,c) = \strat'(q,c)$
unless $c = c_{\M}(1-F_{\strat}(q))$.
\end{proof}

\section{Proofs from Section~\ref{s-wta-approx}}
\label{s-hetero-proofs}

This section contains the proofs comparing the 
expectation of the maximum quality elicited by the
winner-take-all mechanism to that of the optimal mechanism.
Recall that the overall structure of the proof divides
the participants in the optimal mechanism into two
subpopulations---those with cost greater than $\frac{V}{2}$
and those with cost less than or equal to $\frac{V}{2}$. The
following lemmas treat those sub-populations separately.

\begin{lemma} 
% \label{l-comp1}
$\wta \geq \frac{1}{2}E[\max\{q_i~|~i \in \eopt; c_i \leq \frac{V}{2} \}]$. 
\end{lemma}

\proof
First, note that $E[\max\{q_i~|~i \in \eopt; c_i \leq \frac{V}{2} \}]\leq E[\max\{q_i~|~c_i \leq \frac{V}{2} \}]$, so it is sufficient to show that $\wta \geq E[\max\{q_i~|~c_i \leq \frac{V}{2} \}]$. This allows us to reason only about strategic behavior in the winner-take-all mechanism, instead of needing to also understand participation choices in the optimal mechanism. 
We will use the sub-equilibrium lemma to relate equilibrium outcomes in $\WTA$ to the expected output in a particular sub-equilibrium, identified below, that can be compared against $E[\max\{q_i~|~c_i \leq \frac{V}{2} \}]$.

Consider $n-1$ random draws of $(q_i,c_i)$ from $F$, and let $\mu$ denote the {\em median} of the random variable $\max\{q_i~|~c_i \leq \frac{V}{2} \}$. That is, $\mu$ is such that: 
\BEQ
\label{e-mu}
\Pr(\max\{q_i~|~c_i \leq \frac{V}{2}\}> \mu) ~=~\Pr(\max \{q_i~|~c_i \leq \frac{V}{2}\}\leq \mu). 
\EEQ

Consider the strategy profile $\sstrati$ where an agent $i$ participates in $\WTA$ if and only if $q_i \geq \mu$ and $c_i \leq \frac{V}{2}$. We claim that $\sstrati$ constitutes a sub-equilibrium of $\WTA$: by definition of $\mu$, if  $i$ participates when playing $\sstrati$, her expected payoff in $\WTA$ is 
\[ 
E[u_i] = V\cdot\Pr(q_i \geq \max_{j \neq i} q_j~|~j \in \spart(\sstrati)) - c ~\geq~ V\cdot\frac{1}{2} - \frac{V}{2} \geq 0,  
\]  
so that it is individually rational for each agent to participate if all other agents play according to $\sstrati$. 

%Let $\spart$ denote the set of the types of agents who participate in $\WTA$ under the sub-equilibrium strategy profile $\sstrati$: 
%\[ 
%\sparti = \{(a,c)~|~ a_i \geq \mu; c_i \leq \frac{V}{2}\}. 
%\] 
We now analyze the expected outcome in this sub-equilibrium $\sstrati$ of $\WTA$: 
%with $E[\max_n\{a~|c_i \leq \frac{V}{2}\}]$ to obtain the lemma, which we do below:  
\BEAS
E[\max_n\{q_i~|~q_i \geq \mu; c_i \leq \tfrac{V}{2}\}] &=& \int_0^\infty \Pr( \max\{q_i~|~q_i \geq \mu; c_i \leq \tfrac{V}{2}\} > t) dt\\
&=& \int_0^\mu \Pr( \exists q_i~|~q_i \geq \mu; c_i \leq \tfrac{V}{2}\}) dt +  \int_\mu ^\infty \Pr( \max\{q_i~|c_i \leq \tfrac{V}{2}\} > t) dt\\
&=& \mu \Pr(\max \{q_i~|~c_i \leq \tfrac{V}{2}\}  \geq \mu) + \int_\mu ^\infty \Pr( \max\{q_i~|c_i \leq \tfrac{V}{2}\} > t) dt\\
&\geq& \mu \cdot\frac{1}{2} + \frac{1}{2}\int_\mu ^\infty \Pr( \max\{q_i~|c_i \leq \tfrac{V}{2}\} > t) dt\\
&=& \frac{1}{2}\left( \int_0^\mu 1\cdot dt + \int_\mu ^\infty \Pr( \max\{q_i~|c_i \leq \tfrac{V}{2}\} > t) dt\right)\\
&\geq& \frac{1}{2}\left( \int_0^\mu \Pr(\max\{q_i~|c_i \leq \tfrac{V}{2}\} > t) dt + \int_\mu ^\infty \Pr( \max\{q_i~|c_i \leq \tfrac{V}{2}\} > t) dt\right)\\
&=& \frac{1}{2} E\max\{q_i~|c_i \leq \tfrac{V}{2}\}], 
\EEAS
where the inequality in the fourth line follows from the definition of $\mu$ in (\ref{e-mu}), and because $\mu$ depends on the max of $n-1$ random draws of $q_i$ while the max here is over $n$ draws from $q_i$.

By the corollary to the sub-equilibrium lemma (Corollary \ref{c-se-max}), the expected maximum output in any equilibrium of $\WTA$ is at least as large as the expected maximum output in a sub-equilibrium, and therefore specifically in the sub-equilibrium $\sstrati$. Combining this with the inequality just derived, we have  
% \BEAS
\[
\wta \geq E[\max_n\{q_i~|~q_i \geq \mu; c_i \leq \tfrac{V}{2}\}] 
 \geq \frac{1}{2} E[\max_n\{q_i~|c_i \leq \tfrac{V}{2}\}] 
\geq \frac{1}{2}E[\max\{q_i~|~i \in \eopt; c_i \leq \tfrac{V}{2} \}],
\]
% \EEAS
which proves the lemma. 
\endproof 

\begin{lemma} 
% \label{l-comp2}
$\wta \geq E[\max\{q_i~|~i \in \eopt; c_i > \frac{V}{2} \}]$. 
\end{lemma}
\proof
To prove this inequality, we use the sub-equilibrium lemma (Lemma \ref{l-se}) again, albeit with a different strategy profile. Consider the equilibrium $\estrat(\M^*)$ of an optimal mechanism $\M^*$, and let $\eopt$ denote the set of participants in $\M^*$ when agents play according to $\estrat(\M^*)$ as before. Consider a strategy $\sstratj$ for $\WTA$ where an agent participates if and only if she has cost $c > \frac{V}{2}$, and would also participate according to $\estrat(\M^*)$. We claim that $\sstratj$ is a sub-equilibrium of the winner-take-all mechanism $\WTA$, \ie, that the strategy profile where all agents in 
$
\spartj= \{i~|~i\in \eopt; c_i > \frac{V}{2}\}
$ 
participate and all remaining agents do not participate leads to non-negative payoffs for all agents in the {\em winner-take-all} mechanism. 

To show this, consider the following {\em lottery over simple} mechanisms defined by the rewards in the mechanism $\M^*\Mv$: the lottery runs the simple contest which awards equal prizes of $\frac{V}{j}$ each to the top $j$ contestants with probability
\[ 
\Pr(j) = \tfrac{j}{V} (v_j - v_{j+1}). 
\] 
Note that (i) $\sum_{j=1}^n \Pr(j) = 1$ since $\sum_{j=1}^n v_j = V$
(this is derived in Equation~\eqref{e-wsum} in \S \ref{s-homo} for {\em any} feasible mechanism) % from our calculation (\ref{e-psum}) in \S \ref{s-homo}, 
so that the distribution $\Pr(j)$ is a valid probability distribution, and (ii) for any draw of the $n$ values of $(q_i,c_i)$, the expected payoff to the agent with the $j^{\mathrm{th}}$ highest quality is exactly equal to $v_j$ in this lottery over simple mechanisms. That is, this lottery over simple mechanisms is payoff-equivalent (in expectation)  for all agents to the optimal mechanism $\M^*\Mv$.

The expected payoff of an agent in $\eopt$ (\ie, a participating agent) in the {\em optimal} mechanism $\M^*\Mv$ can therefore be written as
\[ 
E[u_i] = \sum_{j = 1}^n \frac{j(v_j - v_{j+1})}{V} \left (\Pr(i \mbox{ belongs to the top } j \mbox{ agents in } \eopt) \cdot \tfrac{V}{j} - c_i \right). 
\] 
Since $\estrat$ is an equilibrium of $\M^*$, every agent who participates must receive non-negative expected payoff. Specifically, consider the agents in $\eopt$ with cost greater than $V/2$. All but the term corresponding to $j = 1$ in the summation are negative for such an agent, since $\frac{V}{j} \leq \frac{V}{2} < c_i$ for $j \geq 2$. So the term corresponding to $j = 1$ must be positive. But for $j=1$, the expression 
$
\left[ \Pr(i \mbox{ belongs to the top } j \mbox{ agents in } \eopt) \cdot \tfrac{V}{j} - c_i \right]
$
is exactly the expected payoff of an agent in $\eopt$ in the {\em winner-take-all} mechanism $\WTA$, if all agents in $\eopt$ participate and all agents not in $\eopt$ do not. Since the expected payoff to this agent when only agents in $\eopt \cap \{i~|~c_i > \frac{V}{2}\}$ participate can only be larger than this value (she is more likely to be the highest-quality agent when competing with a smaller subset of agents), this means that the strategy profile $\sstratj$ where all agents in $\eopt$ with $c_i > \frac{V}{2}$ participate, and the remaining agents do not, is a sub-equilibrium of the winner-take-all mechanism. 

Applying Corollary~\ref{c-se-max} to this strategy profile $\sstratj$, we have 
$
\wta \geq E[\max\{q_i~|~i \in \eopt; c_i > \frac{V}{2}\}], 
$ 
which proves the lemma. 
\endproof

\end{document}